\documentclass[a4paper, 11pt]{article}
 
\usepackage{lipsum}

\usepackage[T1]{fontenc}
\usepackage{tgtermes}
\usepackage{booktabs}
\usepackage{amsmath,amssymb}

\usepackage{natbib}
\bibpunct{(}{)}{;}{a}{}{,} 
\usepackage{enumerate}

\usepackage[british]{babel}
\usepackage{endnotes}

\usepackage[disable]{todonotes}
\usepackage[top=1.75cm, bottom=1.75cm, outer=2cm, inner=2cm, heightrounded, marginparwidth=3cm, marginparsep=0.5cm]{geometry}

\usepackage{type1cm}
\usepackage[T1]{fontenc}
\usepackage{ae,aecompl}
\usepackage{courier}
\usepackage{sectsty}
\allsectionsfont{\usefont{OT1}{phv}{bc}{n}\selectfont}

\usepackage[small,bf,up]{caption}

\def\arxivprefixesep{:}

\newcommand{\eprint}[2][]{%
{\tt\if!#1!#2\else#1\arxivprefixesep\ignorespaces#2\fi}%
}

\usepackage{graphicx}
\usepackage{psfrag}
\usepackage{xspace}
\usepackage{wrapfig}

\newcommand{\hcoplus}{{HCO$^+$}}
\newcommand{\ntwohplus}{N$_2$H$^+$}
\newcommand{\ntwodplus}{N$_2$D$^+$}
\newcommand{\nhtwod}{NH$_2$D}
\newcommand{\htwodplus}{H$_2$D$^+$}
\newcommand{\hthreeplus}{H$_3^+$}
\newcommand{\HII}{H{\scriptsize II}}

\newcommand{\UCHIIs}{UCH{\scriptsize II}s}
\newcommand{\HCHII}{HCH{\scriptsize II}}
\newcommand{\HCHIIs}{HCH{\scriptsize II}s}
\newcommand{\simgt}{\lower.5ex\hbox{$\;\buildrel>\over\sim\;$}}
\newcommand{\simlt}{\lower.5ex\hbox{$\;\buildrel<\over\sim\;$}}
\newcommand{\farcs}{$.\!\!^{\prime\prime}$}

\newcommand{\ga}{\simgt}

\newcommand{\htwo}{H$_2$}
\newcommand{\water}{H$_2$O}

\newcommand{\xco}{$X_{\rm CO}$}

\newcommand{\fco}{F$_{CO}$}
\newcommand{\kms}{km\,s$^{-1}$}
\newcommand{\jykms}{Jy\,km\,s$^{-1}$}

\newcommand{\msun}{$M_\odot$}

\makeatletter
\let\@fnsymbol\@arabic
\makeatother

\begin{document}
\title{The Science Case for ALMA Band 2 and Band 2+3} 
\author{G.~A. Fuller
\endnote{Jodrell Bank Centre for Astrophysics \& UK ALMA Regional Centre Node,
  School of Physics and Astronomy, The University of Manchester, Oxford Road,
  Manchester, M13 9PL, UK}
\and 
  A. Avison$^1$ 
\and
  M. Beltr\'an\endnote{INAF-Osservatorio Astrofisico di Arcetri, Largo
    E. Fermi 5, 50125 Firenze, Italy}
\and 
V. Casasola\endnote{INAF - IRA \& Italian ALMA Regional Centre
Via P. Gobetti 101, 40129 Bologna, Italy} 
\and
P. Caselli\endnote{Max-Planck-Institut f\"ur extraterrestrische Physik
Giessenbachstrasse 1, 85748 Garching, Germany} 
\and
  C. Cicone\endnote{Cavendish Laboratory, University of Cambridge, 19
    J.J. Thomson Avenue, Cambridge CB3 0HE,UK; Kavli Institute for Cosmology,
    University of Cambridge, Madingley Road, Cambridge CB3 0HA, UK }
\and
  F. Costagliola\endnote{Instituto de Astrof\'isica de Andaluc\'ia, Glorieta de
    la Astronom\'ia s/n, Granada, 18008, Spain}
\and
C. De Breuck\endnote{European Southern Observatory (ESO),
  Karl-Schwarzschild-Str. 2, 85748, Garching, Germany} 
\and
  L. Hunt$^2$
\and
  I. Jimenez-Serra$^7$ 
\and
R. Laing$^{7}$ 
\and
  S. Longmore\endnote{Astrophysics Research Institute, Liverpool John Moores
    University, 146 Brownlow Hill, Liverpool L3 5RF, UK}
\and
 M. Massardi$^3$
\and
T. Mroczkowski$^{7}$ 
\and 
R. Paladino$^3$
\and
  S. Ramstedt\endnote{Department of Physics and Astronomy, Uppsala University, Box 516, 751 20, Uppsala, Sweden}
\and 
A. Richards$^1$ 
\and
L. Testi$^{2,7,}$\endnote{Excellence Cluster Universe, Boltzman str. 2, D-85748 Garching bei Muenchen, Germany}
\and 
D. Vergani\endnote{INAF Osservatorio Astronomico di Bologna, via C. Ranzani 1,
  40127, Bologna, Italy}
\and
S. Viti\endnote{Department of Physics and Astronomy, University College London, WC1E 6BT London, UK}
\and
  J. Wagg\endnote{Square Kilometre Array Organisation, Jodrell Bank
    Observatory, Lower Withington Macclesfield Cheshire, SK11 9DL, UK}
}

\maketitle

\begin{abstract}
We discuss the science drivers for ALMA Band 2 which spans the frequency range
from 67 to 90\,GHz.  The key science in this frequency range are the study of
the deuterated molecules in cold, dense, quiescent gas and the study of
redshifted emission from galaxies in CO and other species.  However, Band 2
has a range of other applications which are also presented.  The science
enabled by a single receiver system which would combine ALMA Bands 2 and 3
covering the frequency range 67 to 116\,GHz, as well as the possible doubling
of the IF bandwidth of ALMA to 16\,GHz, are also considered.
\end{abstract}

\listoftodos

\section{Introduction}

ALMA Band 2 spans from $\sim67$\,GHz (bounded by an opaque line complex
of ozone lines) up to 90\,GHz which overlaps with the lower frequency
end of ALMA Band 3.  Band 2 is the only remaining frequency band on
ALMA for which receivers have not yet been designed and
developed. Below we lay out the compelling and varied science which
ALMA Band 2 enables.

Receiver technology has advanced since the original definition of the
ALMA frequency bands. It is now feasible to produce a single receiver
which could cover the whole frequency range from 67\,GHz to 116\,GHz,
encompassing Band 2 and Band 3 in a single receiver cartridge, a so
called Band 2+3 system. In addition, there are now foreseen upgrades
to ALMA system which could double it backend bandwidth to 16\,GHz.  The
science drivers discussed below therefore also discuss the advantages
of these two enhancements over the originally foreseen Band 2 system.

\section*{Level 1 Science Projects}

There are two top level science drivers for ALMA Band 2. The first is the
study of gas in external galaxies where Band 2 makes it possible to study
redshifted CO for both redshift determination and to accurately measure the
cool molecular gas mass.  Band 2 will also allow the study of the properties
and evolution of the dense gas (via the dense gas tracers HCN, HNC and
\hcoplus) in the crucial redshift range where the cosmic star-formation
density is rapidly declining.  The second top level driver is the study of
lowest energy transitions of the simple deuterated molecules which trace the
coldest, densest and more quiescent molecular gas in a range of environments.

\section{Extragalactic Science}

\subsection{Redshift searches \label{sec:redshift}}

Redshifts are essential for understanding the distribution and evolution
of galaxies through cosmic time. ALMA's power as a redshift engine has already
been demonstrated with its serendipitous measurements of redshifts ($z$=4.4)
in just 2 minutes of observing time \citep{2012MNRAS.427.1066S}.  Using
molecular and atomic lines such as CO, [CI], [CII] and \water, ALMA is already
breaking the decades-long monopoly of optical telescopes for redshift
determinations. With full frequency scans to identify lines in mm spectra
becoming the favoured option for redshift determinations
\citep{2009ApJ...705L..45W}, the detection of two transitions of CO provides
an unambiguous redshift.

\begin{figure}[!ht]
\vspace{0pt}
\hbox{
\includegraphics[width=0.48\linewidth]{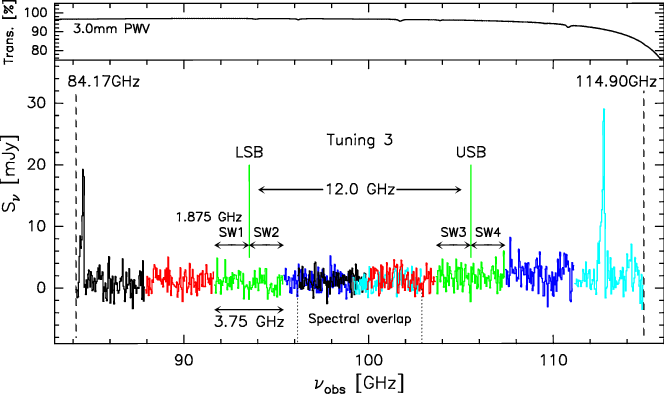}
\hfill
\includegraphics[width=0.48\linewidth]{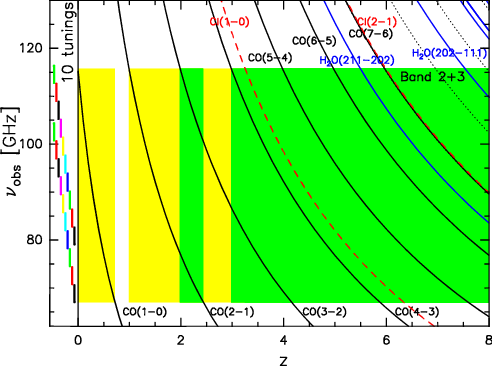}
}
\caption{\footnotesize {\it Left:} Spectral setup used for a redshift search
  using Band 3 (Wei\ss\ et al. 2013). In each tuning, four spectral windows
  covering 1.875\,GHz each were placed in contiguous pairs in the upper and
  lower sidebands. Note that the range 96.2--102.8\,GHz (dotted lines) is
  covered twice.  {\it Right:} Spectral coverage of the CO, [CI] and H$_2$O
  lines as a function of redshift in a combined Band 2+3 receiver. The green
  shaded region marks the redshifts where two or more lines provide an
  unambiguous redshift, while the yellow region marks the redshift range where
  only one line is detectable. Note that there is only a small ``redshift
  desert'' at 0.72$<$$z$$<$0.99. The 10 frequency tunings needed to cover the
  full Band 2+3 frequency range with the current spectral coverage are shown
  in the left panel. With a 16\,GHz contiguous bandwidth, only three tunings
  would be necessary with a two sideband configuration (similar to the current
  Band 3 system) but $8$\,GHz wide sidebands would require four tunings.}
\label{fig:zsearch}
\end{figure}

With the most consistent, and low, atmospheric transparency
(Fig.~\ref{fig:zsearch}, {\it left}), ALMA Band 3, which covers the lowest
excitation CO lines is currently the favoured band for redshift
determinations.  However, Band 3 does not contain any lines in the ``redshift
desert'' ranges 0.37$<z<$0.99 and 1.74$<z<$2.00.  In addition, only at $z>$3
does Band 3 include two CO lines needed to derive an unambiguous
redshift. Confirmation of lower redshifts therefore requires an additional
follow-up observation of a higher frequency line in another ALMA band.

The extension of the low-frequency edge of Band 3 to 67~GHz with a new Band
2+3 receiver would solve most of the limitations of the current Band 3 for
redshift searches. As shown in Fig.~\ref{fig:zsearch}, only a small redshift
desert at 0.72$<$$z$$<$0.99 would remain which would require searches at Band
4 or 5.  The main gains of a Band 2+3 system are that it will allow the
detection of at least 1 line for the entire $z$$>$1 range, and unambiguous
redshifts for $z$$>$2, with the exception of 2.5$<z<$3.0. This is well matched
to the redshift distribution of dusty galaxies, which increases steeply up to
$z$$\sim$2.5 \citep[e.g.][]{2005ApJ...622..772C,2013ApJ...767...88W}. The
success rate of a Band 2+3 redshift survey will thus approach 100\%, a major
revolution compared to current optical redshift surveys which have a
$\sim$50\% spectroscopic completeness rate, and to Band 3 only surveys, where
the completion rate is of order of 70\% \citep{2013ApJ...767...88W}.

The width of the spectral bands is an additional important consideration for
redshift searches, where contiguous spectral coverage is essential. As shown
in Fig.~\ref{fig:zsearch}, {\it left}), the effective 1.875\,GHz bandwidth of
the current basebands requires more than two tunings to fill the 8.0 GHz gap
between the LSB and USB, creating an overlap region in the middle of the
contiguous spectral coverage, but lower coverage at the outer edges where
unfortunately the atmospheric transmission is worst. If the two basebands
could cover at least the full 4.0\,GHz within the sidebands, or cover a
contiguous 16.0 GHz without a gap, this would make both redshift searches (and
spectral surveys in general) more efficient: with the current
2$\times$3.75\,GHz coverage, 10 tunings are required ( Fig.~\ref{fig:zsearch},
{\it right}); with a 2$\times$4.0\,GHz coverage, this can be reduced to 6
tunings; with a contiguous 16\,GHz coverage per tuning, the entire Band 2+3
(except 1.0 GHz at the edges) can be covered with only 3 tunings. This would
increase to 4 tunings for a two sideband configuration (similar to the current
Band 3 system) with 8\,GHz in each sideband.
It is important to note that this high gain in efficiency can {\em only} be
obtained with a combined Band 2+3 receiver and {\bf not} with separate Band 2
and Band 3 receivers.

\subsection{Running out of gas in the third quarter \label{sec:lowJ}}
\label{sec:xco}

The epoch of galaxy formation is most commonly traced by
the cosmic star-formation rate density of the Universe (SFRD).
The SFRD peaks at $z\sim1-3$, during a period usually associated with
the main epoch of galaxy assembly.  In this epoch, roughly half of the
stars present in galaxies today are formed
\citep[e.g.][]{2008ApJS..175...48R,2011ARA&A..49..525S}.  The SFRD
then declines dramatically toward lower redshift, by roughly an order
of magnitude, from $z\sim2$ to $z\sim0$, with the most significant
decrease at $z<1$.

One of the most important diagnostics of the epoch of galaxy assembly
and the successive decline of the SFRD are observations of cool gas.
Theory predicts and observations confirm that the molecular gas fraction
increases with lookback time, by as much as a factor of 7 from $z=0$ to
$z\sim2$ \citep[e.g.][]{2008ApJ...680..246T,2010MNRAS.407.2091G,
  2012MNRAS.426.1178N,2013ApJ...768...74T}.  At $z\sim2$, molecular gas can
contribute as much as 60$-$70\% to the total baryonic inventory.  Such a high
gas content
and the decline in gas mass fraction toward lower redshifts is undoubtedly
associated with the peak of the SFRD, and its decrease toward the present day.
The epoch from $z<1$ is thus crucial to understand how and why galaxies ``ran
out of fuel'' for star formation ``in the third quarter'' of the age of the
Universe.

Recent years have seen the number of high-$z$ galaxies with molecular
gas-mass estimates increase exponentially. However, uncertainties in two
critical factors still hamper the interpretation of observations: one is the
large uncertainty in \xco, the factor that converts observed CO column
densities to \htwo\ masses.
\xco\ for high-SFR galaxies such as luminous infrared galaxies
(LIRGs), ultra-LIRGs (ULIRGs), and sub-millimeter galaxies (SMGs),
typically forming stars in compact dense regions, has been found to be
$\sim$5 times lower than \xco\ for ``Milky-Way-like'' galaxies,
similar in mass but with star formation in more extended disks.  This
results in a bimodality of gas scaling relations such as the
Schmidt-Kennicutt law which relates SFR and gas surface density
\citep[e.g.][]{
  2008ApJ...673L..21D,2010ApJ...714L.118D,2010Natur.463..781T,2013ApJ...768...74T}.
\xco\ is also a strong function of metallicity
\citep[e.g.][]{1988ApJ...325..389M,2011MNRAS.412..337G,2012ApJ...746...69G,2013ARA&A..51..207B}.
Ultimately, assumptions are made based on observations in a single CO
transition, and the more varied galaxy populations and different
physical conditions at high redshift make \xco\ uncertain.

The other uncertainty is linked with the necessity, up to now, of
observing high-redshift galaxies in high-J CO transitions.  Most of
the observations of gas in high-$z$ galaxies have been carried out in
CO(3-2) or higher-J lines
\citep[e.g.][]{2008ApJ...673L..21D,2010ApJ...714L.118D,2010MNRAS.407.2091G,2013ApJ...768...74T}.
However, the estimate of gas masses is ultimately based on the CO(1-0)
transition, so that ratios of the higher-J lines relative to
1$\rightarrow$0 must be assumed.  As illustrated in the right panel of
Fig. \ref{fig:bands_sled}, these ratios are critically related to
excitation conditions, which are known to vary between low-
vs. high-SFR galaxies, and compact vs. extended star-forming regions
\citep[e.g.][]{2005A&A...440L..45W,2011MNRAS.410.1687D}.  Observing
only high-J lines skews mass estimates toward the warm, dense gas
traced by these transitions, and can cause molecular gas masses to be
severely underestimated \citep[e.g.][]{2009ApJ...698L.178D,2010ApJ...718..177A,2011MNRAS.412.1913I}.

\subsubsection{Band 2 and cool gas mass }

\begin{figure}[!ht]
\vspace{0pt}
\hbox{
\includegraphics[width=0.48\linewidth]{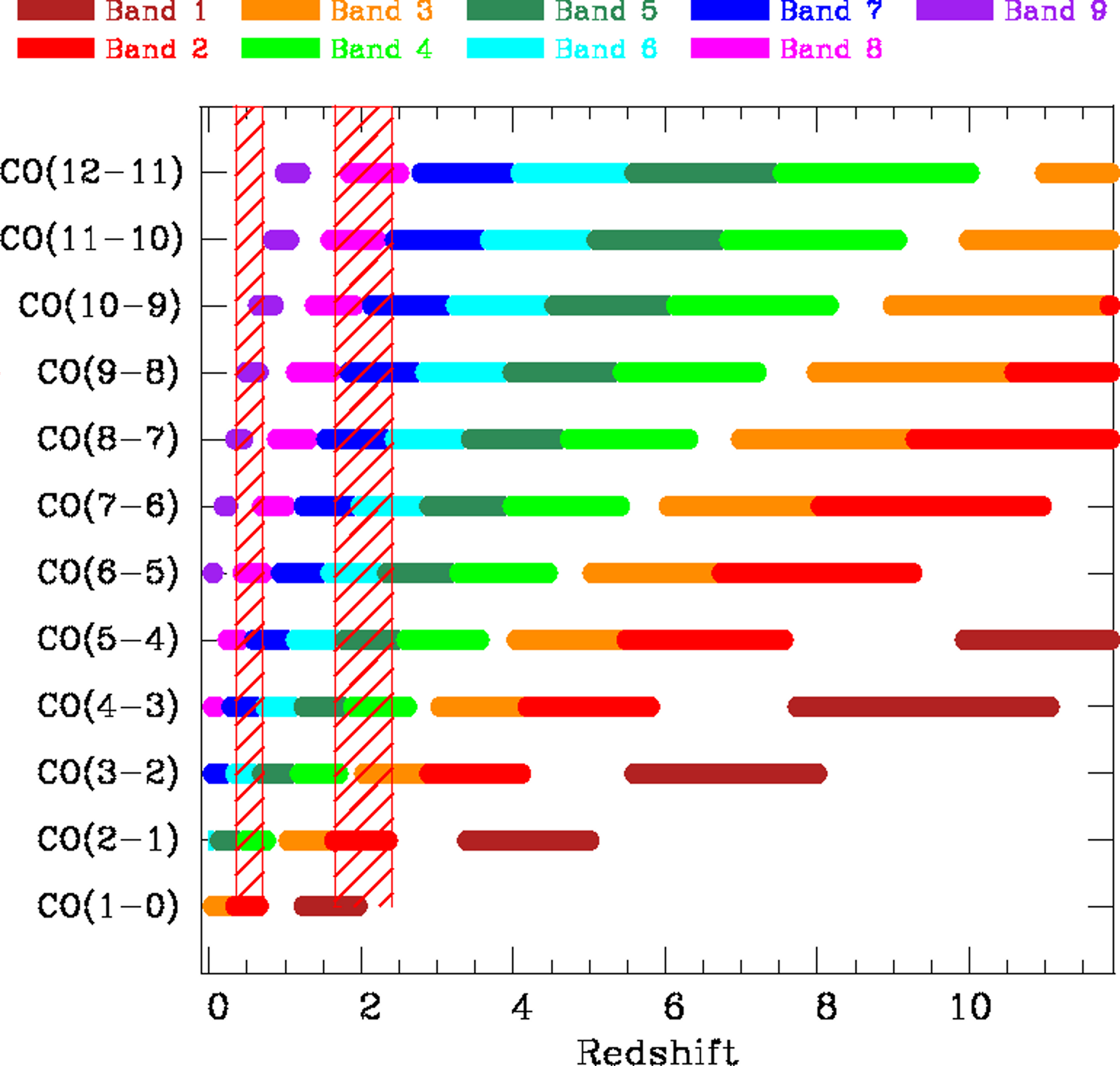}
\hfill
\includegraphics[width=0.48\linewidth]{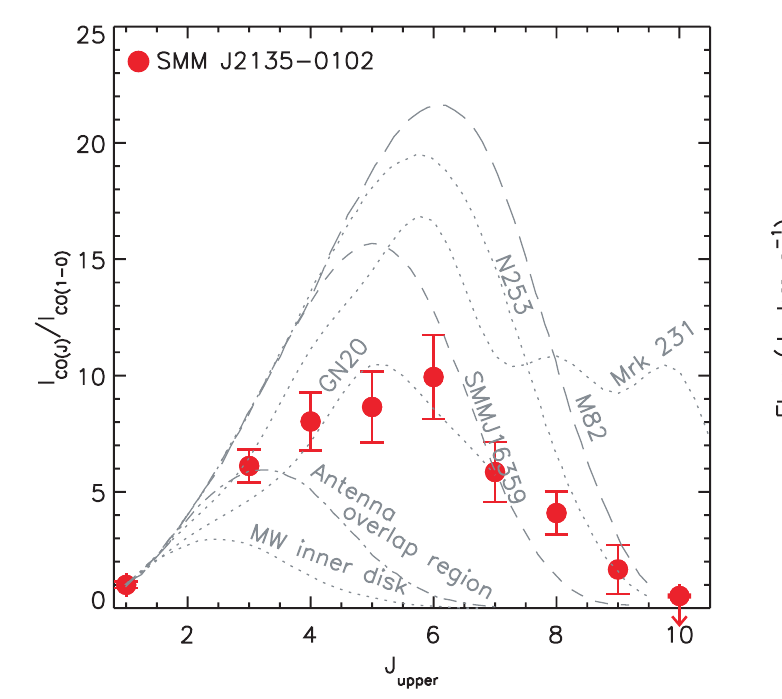}
}
\caption{\footnotesize{ \it Left:} CO ladder coverage by the ALMA receivers plotted against redshift. 
The vertical hatched strips show the potential of ALMA with Band 2
to measure multiple transitions in two crucial redshift regions,
the first where the Universe is 3/4 of its present age (the ``last quarter''),
and the second where the cosmic SFRD peaks.
{\it Right:} Integrated $^{12}$CO SLED for a SMG,
SMM J2135-0102 at $z=2.3$, taken from \citet{2011MNRAS.410.1687D}.
The gray curves show SLEDs from different galaxies, illustrating
the contrast between starbursts such as NGC\,253 and M\,82, 
and more quiescent systems such as the Milky Way inner disk.}
\label{fig:bands_sled}
\end{figure}

\textit{Band 2 can mitigate, if not completely resolve, these uncertainties}
while also constraining the unknown physical conditions in these high-z
galaxies.  CO(1-0) is unobservable at $z>0.3$ with only Band 3, and CO(2-1) is
recovered by Band 3 only at $z\sim$1.0.  However, Band 2 enables CO(1-0)
measurements over this crucial range in redshift, from $z\sim0.7$ to
$z\sim0.3$ where SFRD falls dramatically and the Universe is about 3/4 of its
present age ($0.3<z<1.0$ corresponds to lookback times from 3.4\,Gyr to
7.7\,Gyr).
Figure \ref{fig:bands_sled} (left panel) illustrates the CO ladder and
the band coverage as a function of redshift.  From $z=0.29$ to
$z=0.72$, Band 2 combined with the other ALMA bands enables
observations of between seven and nine transitions, including the four with
lowest J; Band 2 covers 1$\rightarrow$0, thus making possible a
complete assessment of excitation conditions and the CO spectral line
energy distribution (SLED).
 
No CO lines can be observed in Band 1 until $z\sim1.2$, where the 1-0
transition enters the band; Band 2 recovers CO(2-1) at $z\sim1.6$ which
remains observable with Band 2 until $z\sim2.4$.  Thus, in the redshift range
where the SFRD is peaking, from $z\sim1.6$ to $z\sim2.4$, Band 2, together
with the other ALMA bands, provides coverage of between eight and eleven CO
transitions; Bands 2+1 measure the two lowest-J transitions, indispensable for
accurate gas mass estimates.  As shown in Fig. \ref{fig:bands_sled} (right
panel), such band coverage is fundamental because of the shape of the CO
cooling curve, which peaks around 6$\rightarrow$5 in the case of starbursts,
or toward lower J in the case of more quiescent disks.  The factor to scale
the higher-J lines to 1$\rightarrow$0 depends on excitation, and is virtually
impossible to estimate without multiple transitions.  Without the two lowest
CO transitions, the mass in molecular gas can be underestimated by more than a
factor of 2, depending on the excitation of the gas
\citep{2009ApJ...698L.178D}.

Expanding the frequency coverage of ALMA by installing Band 2 will
allow the almost complete exploration of
the cool gas content in galaxies in this ``third quarter" of the age
of the Universe.  There would remain only a gap of $\sim$0.7\,Gyr in
the evolution of the
Universe (between $z\sim0.84$ and $z\sim1$, where CO(2-1) shifts out of Band 4
and into Band 3; CO(1-0) is recovered by Band 1 at $z\sim1.2$).
These low-J transitions are necessary to accurately estimate the cool
molecular gas mass, as well as image its distribution and kinematics.
This epoch in redshift is crucial for observationally
constraining the physical mechanisms which quench star 
formation and cause galaxies to transform from blue star-forming systems to 
``red and dead'' ones.

\subsubsection{Band 2 sensitivity and resolution}

The Band 2 specifications ($\sim$12\,\jykms\ in 60\,s integration time with 50
12m antennas) make ALMA an extremely sensitive facility for measuring cool
molecular gas mass at high redshift.  For a gas mass of $10^9$\,\msun\ at
$z=0.7$ (assuming a Galactic conversion factor), we expect a CO(1-0) flux
\fco\ of $\sim$ 0.004\,\jykms\ per channel, assuming 10\,\kms\ channels and a
velocity width $\sim$100\,\kms.  We would obtain a 5$\sigma$ detection in
about 25 minutes.  For $10^8$\,\msun\ of gas (at $z=0.7$ as before), we would
expect a smaller velocity width, say $\sim$80\,\kms; with 20\,\kms\ channels,
we would be able to achieve a 5$\sigma$ detection in 2.5\,hr.  This means that
at these redshifts, in reasonable times, ALMA is able to image and study the
kinematics of the cool gas masses associated with galaxies
$\sim$10$^{8-9.6}$\,\msun.  It is the galaxies in this range of masses which
are evolving the most at these redshifts (in the ``last quarter'' of the age
of the Universe), and constitute an important constraint for galaxy evolution
schemes \citep[e.g.][]{2006ApJ...651..120B,2012MNRAS.426.2797W}.

At higher redshifts, Band 2 probes CO(2-1) from $z=1.6$ to $z=2.4$; at the
same redshifts, CO(1-0) is observable with Band 1.  At $z=2.4$,
$5\times10^9$\,\msun\ of gas, spread over ten 20\,\kms\ channels, would give
\fco$\sim$0.0017\,\jykms\ per channel (again assuming a Galactic conversion
factor).  A 5$\sigma$ detection could be achieved in $\sim$2.1\,hr.  This
redshift range is where the cosmic SFRD peaks, and is where the gas-mass
fractions are among the highest yet explored with current facilities.  By
enabling more accurate gas mass estimates with low-J lines as well as probing
the dynamics of this cool gas, ALMA Band 2 would open up a new era in the study
of galaxy evolution.

In an extended configuration, at $z=0.7$, ALMA would resolve
structures on scales of $\sim$400\,pc. At $z=2.4$, the situation is
not significantly worse, since this same configuration with 0\farcs06
resolution would probe structures on scales of $\sim$500\,pc.  Such
scales are where gas scaling relations start to break down
\citep[e.g.][]{2010A&A...510A..64V}, and are thus a suitable
resolution for our science goals.

\subsection{Dense star-forming gas and molecular outflows in nearby galaxies \label{sec:dense}}
\begin{figure}
\includegraphics[width=.45\columnwidth]{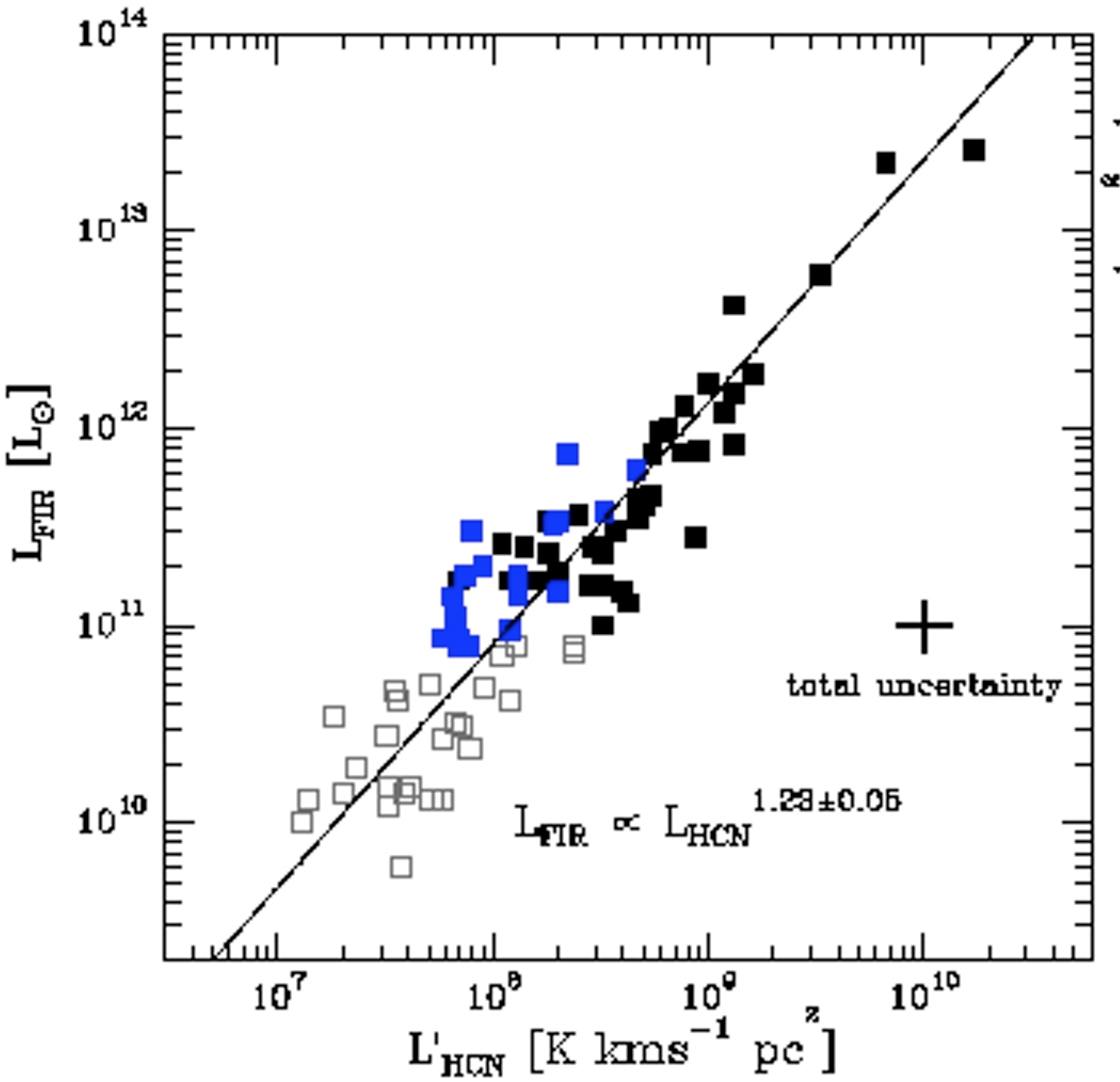}\quad
\includegraphics[width=.6\columnwidth]{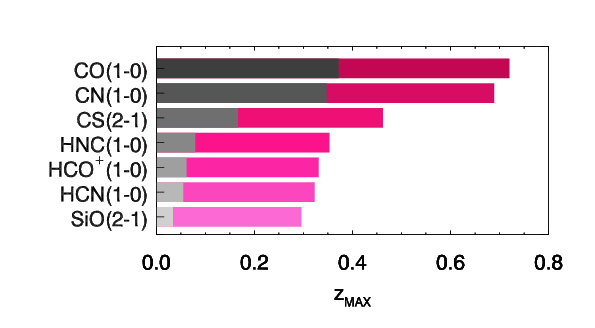}\\
\caption{\footnotesize{\it Left:} Correlation between the star formation rate
  (traced by total IR luminosity) and HCN luminosity in a sample of nearby
  galaxies from Garcia-Burillo et al. (2012) (see also Gao \& Solomon 2004);
  {\it Right:} Molecular transitions of interest for studying the physical
  properties of massive molecular outflows, plotted as a function of the
  maximum redshift at which they are detectable with ALMA.  Grey bars show the
  current situation and pink bars show the improvement given by the Band 2.}
\label{fig:HCN}
\end{figure}

Although observations of CO line emission serve as an effective means
of studying the cold molecular gas reservoir in the interstellar
medium of high-redshift galaxies, probing the dense molecular gas
($n_{H_2} \ga$10$^5$~cm$^{-3}$) more directly associated with
star-formation is best achieved through observations of higher dipole
moment molecules, such as HCN, HNC or HCO$^+$. The luminosity in the
HCN~\textit{J}=1-0 line is known to correlate tightly with infrared
(IR) luminosity in nearby galaxies
\citep[][Fig.~\ref{fig:HCN}]{2004ApJS..152...63G,2004ApJ...606..271G,2012A&A...539A...8G},
a correlation which extends down to the IR luminosities of dense
Galactic molecular cloud cores, $L_{IR} \ga 10^{4.5}~L_{\odot}$ (Wu et
al.\ 2005). For warm gas ($\sim$70~K), the mass in dense gas can be
estimated from the HCN\textit{J}=1-0 line luminosity following,
$M_{dense} \sim 7 \cdot L^{\prime}_{HCN}$~M$_{\odot}$
\citep{2004ApJ...606..271G}.  As this dense gas in our own Galaxy is
generally associated with sites of ongoing star-formation (indicated
by the IR luminosity), it is believed that the presence of strong
HCN~\textit{J}=1-0 line emission in a high-redshift galaxy may be
interpreted as the signature of an ongoing starburst
\citep{2003Natur.426..636S}. By comparing the
dense gas masses estimated from HCN~\textit{J}=1-0 with the total
molecular gas masses estimated from low-J CO line emission, one can
estimate the dense fraction in galaxies over Cosmic time. Similarly,
the dense gas mass can be compared to the star-formation rates
(obscured and un-obscured) in order to estimate a true star-formation
efficiency.

With the sensitivity of current facilities, detection of the low-J
transitions of the dense gas tracers HCN, HCO+ and HNC ($\nu_{rest}
\sim 89$~GHz) in emission at high-redshift (z > 1) is only possible
for the most luminous ($L_{FIR} \ga 10^{12}$~L$_{\odot}$), or
gravitationally lensed objects.  Although more than 100 high-redshift
galaxies have now been detected in CO line emission (see review by
Carilli \& Walter 2013), of these only a few have also been detected
in HCN J=1-0 line emission
\citep[e.g.][]{2003Natur.426..636S,2004ApJ...614L..97V,2005ApJ...618..586C}. The
HCN J=5-4 line has been detected in the strongly lensed quasar,
APM08279+5255 at z=3.9 \citep{2005ApJ...634L..13W}.  Because they are
typically more luminous in the IR than the nearby galaxy sample, these
high-redshift galaxies may be used to constrain the extreme end of the
HCN to IR luminosity correlation ($L_{FIR} \ga
10^{12}$~L$_{\odot}$). Although the current data suggest a deviation
from the correlation at this high luminosity end
\citep{2007ApJ...671L..13R,2012A&A...539A...8G}, this needs to be
confirmed through observations of more objects with comparable
luminosities. HCO$^+$ is also known to be a good tracer of dense gas
in extragalactic environments
\citep[e.g.][]{2006ApJ...640L.135G,2006ApJ...645L..13R}, however
observations of this species are more difficult as it is an molecular
ion \citep{2007ApJ...656..792P}. 

\begin{figure}
\centering
\includegraphics[width=.6\columnwidth]{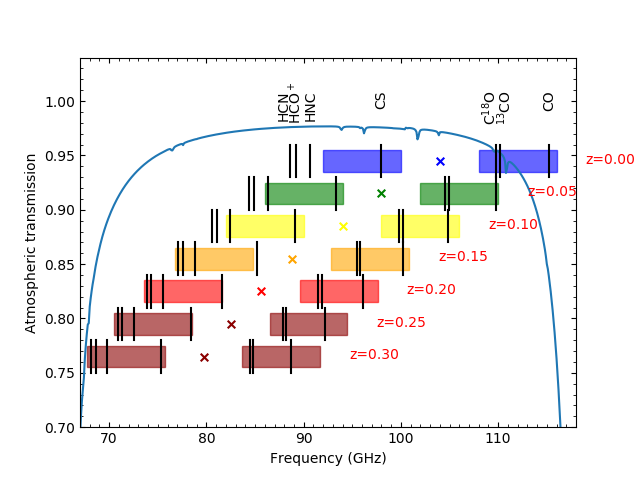}
\caption{\footnotesize Band 2+3 configurations observing CO, and its isotopologues, and dense gas tracers (HCN, HCO$^+$, HNC and CS) as a function of redshift from z=0 to 0.3. }
\label{fig:dense-co-z}
\end{figure}

Figure~\ref{fig:dense-co-z} shows the value of Band 2+3 with a 16\,GHz bandwidth (8\,GHz in each sideband) for simultaneously studying the relative evolution of molecular reservoirs in galaxies (traced by CO J=1-0 and its isotopolgues) and the dense star forming gas (traced by HCN, HCO$^+$, HNC and CS). At redshift z=0, all the J=1-0 transition of CO species plus CS J=2-1 can be observed simultaneously. Moving to higher redshifts additional dense gas tracers can also be observed simultaneously. Beyond z=0.15, the J=1-0 transitions of HCN, HCO$^+$ and HNC can all be observed simultaneously with the CO species while beyond z=0.20 all four dense gas tracers are simultaneously observable with all the CO species. If the IF bandwidth in each sideband was expanded to 9\,GHz, then at least two dense gas tracers (CS and HNC) would be observable with all the CO isotopolgues at all redshifts $0\leq z\leq0.3$. All four dense gas tracers would be simultaneously observable (with the CO species) for all redshifts up to $z=0.3$ if the IF bandwidth was 10\,GHz or greater in each sideband. 

\subsubsection{Molecular outflows and AGN feedback}

The massive molecular outflow in the ULIRG, Mrk\,231, which is the best
studied so far, has been detected not only in low-J CO transitions
\citep{2010A&A...518L.155F,2012A&A...543A..99C}, but also in the high density
molecular tracers HCN(1-0), HCO$^+$(1-0), HNC(1-0)
\citep{2012A&A...537A..44A}.  Moreover, the broad wings that trace the outflow
are \textit{more prominent} in the higher density species (HCN) than in the
low-J CO lines, as indicated by the line core-to-wings intensity ratios which
are up to a factor of $\sim$4 higher in the HCN than in the CO lines.  These
results suggest that not only very dense (${\rm n>10^4 cm^{-3}}$) molecular
gas survives in the outflow, but also that these high dipole moment molecules
are enhanced in the wind, and they may therefore \textit{constitute the best
  tool to explore the outflow}.  A possible explanation for such enhanced HCN
emission in the massive molecular wind can be the presence of strong shocks;
however, the hypothesis of strong shocks seems to be in contrast with the
results of the CO excitation study of the outflowing gas in Mrk 231
\citep{2012A&A...543A..99C}.  Given the direct link between HCN and star
formation rate \citep{2004ApJ...606..271G}, the detection of the outflow in
HCN, HCO$^+$ and HNC is also intriguing in light of the recent models of
triggered by AGN feedback in which star formation occurs within massive
molecular outflows (``positive AGN feedback'',
\citealt{2012MNRAS.427.2998I,2013MNRAS.433.3079Z,2013ApJ...772..112S}).

Other important molecular transitions will come {\it for free} within Band
2(+3), which include CS(2-1) 97.98\,GHz and SiO (2-1) 86.85\,GHz.  CS, being
particularly resistant to shocks and UV photo-dissociation, is a very good and
unbiased tracer of dense molecular gas, and can be used to determine the
amount of dense gas in the outflows and its relationship with the diffuse
component traced by the lowest J transitions of CO.  Silicon monoxide (SiO) is
an unambiguous tracer of strong shocks: these can modify the chemistry of the
local ISM, by destroying dust grains and injecting silicon and SiO into the
gas phase, therefore resulting in an abundance of SiO enhanced by orders of
magnitude. This molecule represents an independent tool to test the presence
of shocks and their relevance in the feedback process.  The J=1-0 transition
of CS and SiO are covered by the ALMA Band 1 and the Q Band of JVLA , but the
J=2-1 transitions are {\it crucial to study the excitation of these species}
in the outflow.
 
Summarising, HCN(1-0), HCO$^+$(1-0) and HNC(1-0) are probably
\textit{privileged} tracers of massive molecular outflows, and can be
used explore the outflow in nearby galaxies, where we have enough
signal and resolution to resolve them. Moreover, with Band 2 we can
exploit CS(2-1) and SiO(2-1) to study in detail the physical
properties of the outflows and the role of shocks, therefore
discriminating between different models of AGN feedback.  Without Band
2, all of these transitions are only detectable in sources at redshift
$z<0.05$, which means that the number of possible targets in which to
trace quasar feedback mechanisms is limited to just a few objects
(e.g. Mrk 231).
 
Such studies would also benefit from increased bandwidth: on
the one hand, the continuum can be estimated and subtracted with
greater accuracy, and on the other, most of the previously mentioned
transitions can be observed simultaneously with a single observation. A
combined Band 2+3 receiver would be more efficient for this work than a
separate Band 2 receiver.


\subsubsection{The prospects of Band 2/2+3 for dense gas tracers}

Band 2(+3) receivers on ALMA open up the possibility of detecting (and
imaging) the \textit{J}=1-0 transitions of HCN, HCO+ and HNC in star-forming
galaxies and AGN out to $z$=0.3, compared to only $z<$0.05 with the current
Band 3 capabilities. \textit{Extending to 0.05$<$z$<$0.3 opens up a volume 185
  times larger to enable detailed case studies of targets that are rare in the
  local Universe, but more representative of the galaxy populations at
  high-$z$}. As the number density of IR luminous star-forming galaxies is
increasing with redshift, surveys of HCN \textit{J}=1-0 with Band 2 would
provide the only means of constraining the apparent non-linearity at the
highest end of the $L_{IR}-L^{\prime}_{HCN}$ relationship. The spatial
resolution achievable with ALMA at these frequencies means that we will probe
the spatial distribution of dense gas in the interstellar medium of galaxies
over the last two billion years of the Universe. In the 0.05$<$z$<$0.3 range,
the Universe evolved by 3 Gyr, while the spatial scale changes from 1$''$/kpc
to 4$''$/kpc. We also note that these lines only enter Band 1 at $z$$>$1,
where similar studies will be limited by sensitivity and spatial resolution.

\subsection{Galaxy environment \label{sec:environment}}

Galaxies occur in a range of environments, from close-pairs to clusters which
are the largest collapsed structures in the Universe with total masses up to
$10^{15}$\,\msun\ \citep[e.g.][]{2009astro2010S...4A}. One of the most
important issues is the spatial and temporal evolution of star formation
activity within these environments.  A key requirement for star formation is
the presence of a reservoir of dense, cold gas that can be efficiently
converted into stars. This is especially crucial for galaxies in rich
clusters, because they are expected to be affected by mechanisms able to
remove cold gas from the haloes and disks of infalling galaxies \citep[e.g.,
ram pressure stripping,][]{1972ApJ...176....1G} or to prevent further cooling
of gas within galaxies' dark matter haloes \citep[starvation or strangulation,
e.g.][]{1980ApJ...237..692L, 2002ApJ...577..651B}. This environmental
dependence profoundly influences the evolutionary histories of galaxy clusters
and the star formation-galaxy density relation represents a well-established
observational hallmark of how galaxies evolve as a function of environment
\citep[e.g.][]{2009MNRAS.395L..62G}.

Studies of star formation activity based on multi-wavelength tracers
have shown a gradual truncation, from $z=0$ to $z\sim$1, in the cores
of rich clusters \citep[e.g.][]{1998ApJ...499..589H, 2001ApJ...547..609E, 2003ApJ...584..210G, 2009ApJ...705L..67P}. Determining the nature and modes of star formation requires a
robust understanding of the relationship between the gas content of a
galaxy and its star formation rate. Remarkable progress has been made
in understanding the conversion mechanisms in field galaxies
\citep[e.g.][]{ 2002ApJ...569..157W, 2008AJ....136.2846B,
  2010ApJ...713..686D, 2010MNRAS.407.2091G,
  2010Natur.463..781T,2011A&A...528A.124C}, but the cold and dense gas
fueling the star formation has been difficult to investigate in
clusters \citep[e.g][]{ 2012ApJ...752...91W,2013A&A...558A..60C},
despite their exceptional opportunities for cosmological studies. So
far, cold gas has been detected in cluster galaxies only up to
$z\sim$0.5, especially in the outskirts of rich clusters
\citep[e.g.][]{2009MNRAS.395L..62G,2013A&A...557A.103J}, while the
range $z\sim0.5-1$ is yet unexplored.

For the first time, ALMA Band 2 will offer the opportunity to study
the CO(1-0) line in group and cluster galaxies at $z\sim0.3-0.7$, in a redshift
range where the gas-star conversion in clusters is completely
unexplored. Although similar redshift ranges are observable in other
ALMA bands,
Band 2 enables accurate estimates of molecular gas content
(see Sect. \ref{sec:lowJ}).
In addition, Band 2 offers a field of view $\sim2-3$ times larger (depending
on the redshift range) than other ALMA bands at higher frequencies; clusters
at $z\simgt$1 are generally $\simlt$1\,arcmin in size
\citep[e.g.,][]{2012ApJ...752...91W,2013A&A...558A..60C} so that the Band 2
field-of-view of $\sim$70-90'' gives coverage of the entire cluster and
resolves individual cluster members with only a single pointing.  Observations
of cluster members covering a wider radial range can help distinguish the
various physical processes expected to play a role (e.g., ram-pressure
stripping, strangulation, tidal interactions, and mergers) because these
processes peak in effectiveness at different clustercentric radii
\citep{2010MNRAS.406.1533D,2007ApJ...671.1503M}. How, when, and where such
mechanisms affect the evolution of galaxies has yet to be explored.



\null
\bigskip
\bigskip
\bigskip
\newpage

\section{Deuterated Molecules: Probing the Coldest, Densest, Most Quiescent
  Gas}
\label{sec:deuterium}

Deuterium is produced by primordial nucleosynthesis and destroyed in stars. In
the local universe this has resulted in an abundance of deuterium of
$\sim10^{-5}$ of that hydrogen \citep{1998SSRv...84..285L}. Nevertheless a range
of deuterium-containing molecules have been detected in interstellar
clouds with abundances approaching 50\% that their hydrogen-containing
counterparts \citep[e.g.][]{1998A&A...338L..43C,2005ApJ...619..379C}
(Figure~\ref{fig_dratios}) and even species containing multiple D atoms such
as ND$_3$ have been detected \citep{2002ApJ...571L..55L}.

The factor $\sim10^5$ enhancement in singly deuterated species compared to
their hydrogen counterparts and up to $10^{13}$ for multiply deuterated \citep[e.g.][]{2017A&A...600A..61H}
species is a result of two factors. The first is the small zero-point energy
difference between the deuterium and hydrogen containing species. This
drives the chemical networks producing these species to favour the
deuterated species at low temperature. At temperatures below $\sim20$~K the
reaction
$$ \textrm{H}_3^++ \textrm{HD}\rightarrow
\textrm{H}_2\textrm{D}^++\textrm{H}_2 + 230 \textrm{K} $$
\citep{1974ApJ...188...35W} enhances the abundance of H$_2$D$^+$ as the back
reaction is suppressed.  The H$_2$D$^+$ formed can subsequently react with CO
and N$_2$ to produce DCO$^+$ and \ntwodplus\ respectively. With the decrease
in abundance of the neutral gas species as they condense on to the surfaces of
grains (freeze-out) in high density, lower temperature regions
\citep[e.g.][]{1999ApJ...523L.165C}, the dominant destruction routes for
\hthreeplus\ and \htwodplus\ become less effective.  This, combined with the
enhanced rate of \htwodplus\ production, leads to increasing levels of
deuteration. Eventually, the reduction in the gas-phase abundance of CO
results in \ntwodplus\ becoming the major product of the destruction of
\htwodplus. Deuterium-containing molecules and ultimately \ntwodplus\ are
therefore highly selective probes of the coldest, densest regions of molecular
gas in a range of objects.

\subsection{Prestellar Cores}

 Prestellar cores are the cold ($T\sim10$~K), dense
($n>10^4$~cm$^{-3}$) and quiescent (dominated by the thermal pressure)
reservoirs in nearby low mass star forming regions, with gravitationally
unstable, centrally concentrated density profiles
\citep[e.g.][]{2008ApJ...683..238K}. \begin{wrapfigure}{r}{8cm}
\vspace*{-0.cm}
\caption{\footnotesize Map of the dust continuum emission at 1.3mm
  towards the prestellar core L1544 (Left). The outer contour
  indicates the region where dark cloud chemistry dominates while 
  the inner contour, which encloses the dust peak, shows the deuteration
  zone. (Right) Schematic showing the main chemical processes in these
  two regions. The species in blue indicate the reaction partners.
  \citep{2012A&ARv..20...56C}. }
\label{fig_l1544}
\centering
\includegraphics[width=8cm]{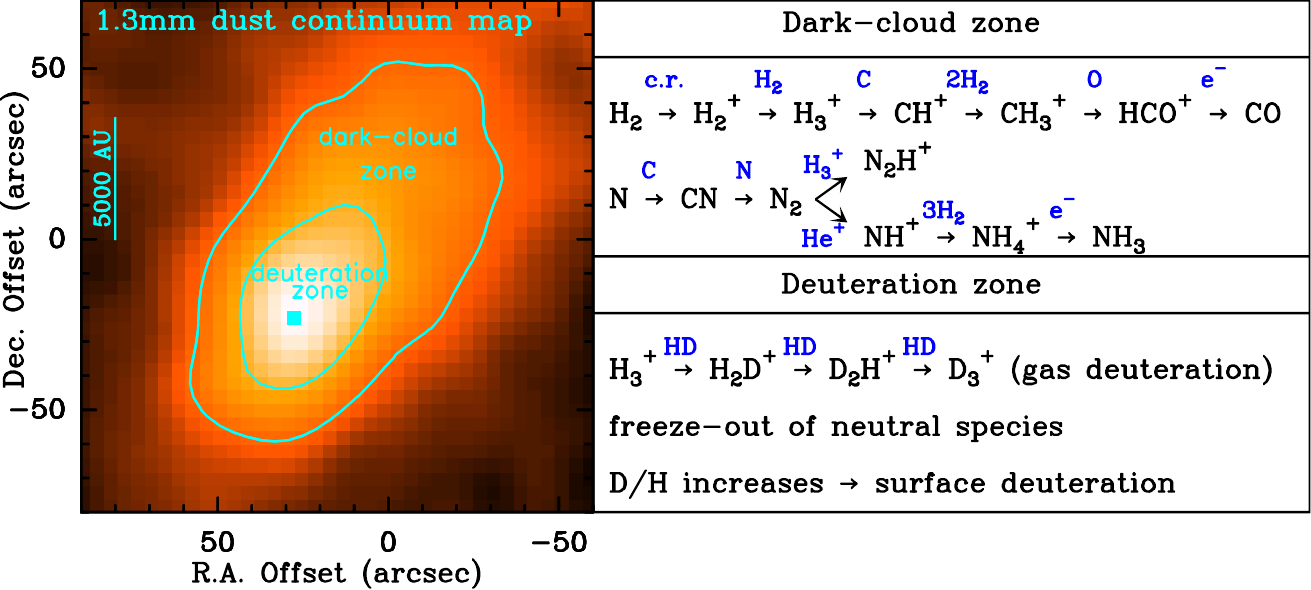}
\end{wrapfigure}
 They show 
the largest deuterium
fractions and highest degree of CO freeze out among the starless cores
\citep{2005ApJ...619..379C}.  These objects are the transitional stage from
molecular cloud to star forming clump and their properties set the initial
conditions for the formation of the next generation of individual low mass
protostars \citep[e.g.][]{1999MNRAS.305..143W}.  The properties of these
future stellar cradles provide crucial tests of star formation theories and
deuterated molecules are key tracers of their physical conditions and dynamics
in their central regions. 

\begin{table}
\caption{\small Deuterated Species and Transitions in ALMA Band 2}
\label{tab_dspecies}
\centering
\begin{tabular}{lcclcc}
\toprule
\multicolumn{6}{c}{Deuterated Species} \\
\cmidrule(r){1-6} 
Molecule &  & Freq. & Molecule &  & Freq. \\
         &            & (GHz)    &      &            & (GHz)\\  \midrule
CH$_2$D$^+$ & 1(1,0)-1(1,1) & 67.273 &  DN$^{13}$C & J=1-0 & 73.367\\
D$^{13}$CO$^+$ & J=1-0 & 70.733&        DNC & J=1-0 & 76.306 \\                
D$^{13}$CN & J=1-0 & 71.175 &           DOC$^{+}$ & J=1-0 & 76.386 \\          
DCO$^+$ & J=1-0 & 72.039&               N$_2$D$^+$ & J=1-0 & 77.108\\             
C$_2$D & N=1-0 & 72.108 &               NH$_2$D & 1(1,1)0 - 1(0,1)0 & 85.926 \\
DCN & J=1-0 & 72.415\\
\bottomrule
\end{tabular}
\end{table}

\begin{wrapfigure}{r}{8cm}
\caption{\footnotesize Schematic physical and chemical structure of prestellar core
  indicating the important model probes \citep{2011IAUS..280...19C}.}
\label{fig_pscmodel}
\includegraphics[width=8cm]{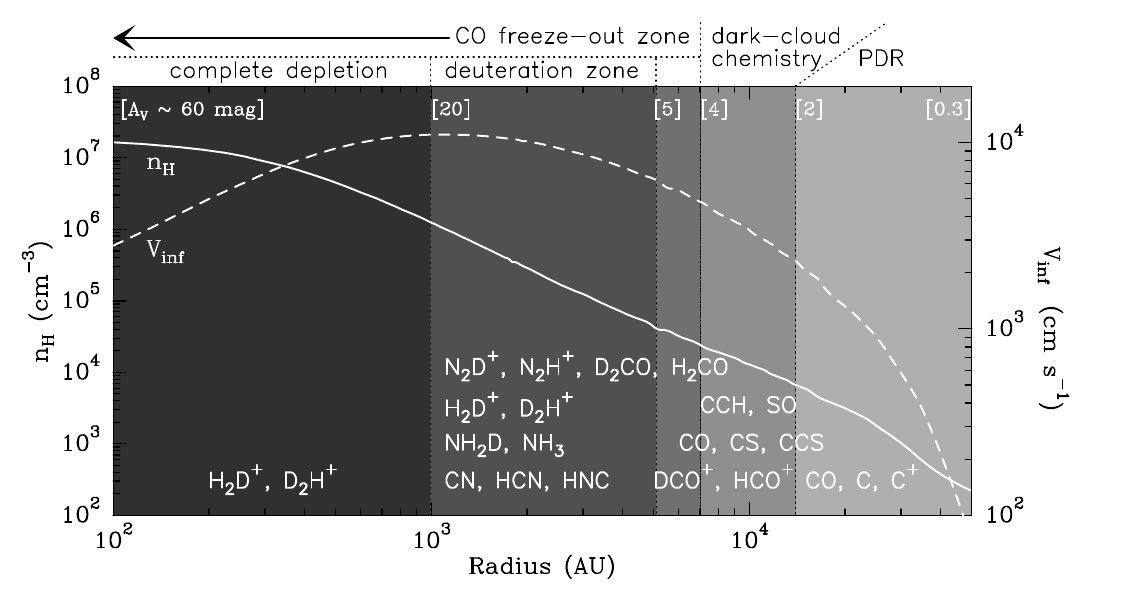}
\end{wrapfigure}

Figure~\ref{fig_l1544} shows the size and location of the deuteration zone
within the prototypical pre-stellar core L1544 in the Taurus molecular cloud, 
with the peak emission in the deuterated species coincident with the dust
continuum peak \citep{2002ApJ...565..344C,2006ApJ...645.1198V}. A schematic of
the structure of a pre-stellar core based on detailed modelling of L1544 is
shown in Figure~\ref{fig_pscmodel} indicating the important tracers of the
different regions of the core \citep{2011IAUS..280...19C}.

In the outer regions of the core, where the density rises above
$\sim10^4$~cm$^{-3}$ a dark-cloud chemistry dominates with species like
HCO$^+$, C$_2$H, H$_2$CO tracing the material, with DCO$^+$ becoming important
as the low temperature enhances the H$_2$D$^+$/H$_3^+$ ratio (see previous
section).  Moving to the higher density interior, where there is less gas
phase CO and the region (in this model) of maximum infall velocity
($\sim0.1$~km~s$^{-1}$), the nitrogen bearing such as \ntwohplus\ and NH$_3$
and then their deuterated counterparts become the dominant tracers. The inner
most regions are currently unexplored. Some authors predict complete
freeze-out of species heavier than He \citep[e.g.][]{2004A&A...418.1035W}, so
that the only likely tracers of the gas at densities above about
10$^6$\,cm$^{-3}$ are the deuterated variants of H$_3^+$, H$_2$D and
D$_2$H$^+$.  However, observations with the Plateau de Bure interferometer
(PdBI) have shown deuterated ammonia strongly peaked at the dust peak
\citep{2007A&A...470..221C}, while water vapour emission from the central
1000\,AU has also been detected \citep{2012ApJ...759L..37C}.  Thus,
easy-to-observe deuterated species such as N$_2$D$^+$ can still be present in
pre-stellar core centres.  As a core evolves to higher average density the
global \ntwodplus\ to \ntwohplus\ ratio can provide an indicator of the
dynamical evolution of the prestellar core \citep{2005ApJ...619..379C}.

The small infall velocities and low degree of turbulence in these
regions \citep{2011IAUS..280...19C} highlights an especially
important feature of the J=1-0 transition of \ntwodplus\ (and
\ntwohplus) compared to their higher J transitions. The Band 2 J=1-0
transitions of these species, unlike the higher J transitions, have an
isolated single hyperfine component. This component provides a
sensitive tracer of the velocity and line width structure of regions
without the ambiguities introduced by optical depth and velocity
dispersion-driven blending of overlapping hyperfine components which
affect the higher J transitions.

\subsection{Protostellar Envelopes: Tracing Protostellar Evolution}

\begin{figure}[ht]
\centering
\begin{minipage}{0.46\textwidth}
  \caption{\footnotesize The correlation of global against source average dust
    temperature. Sources evolve to higher dust temperature as they start to
    heat the circumstellar envelope \citep{2009A&A...493...89E}.}
\label{fig_class0-evol}
\includegraphics[width=7.5cm]{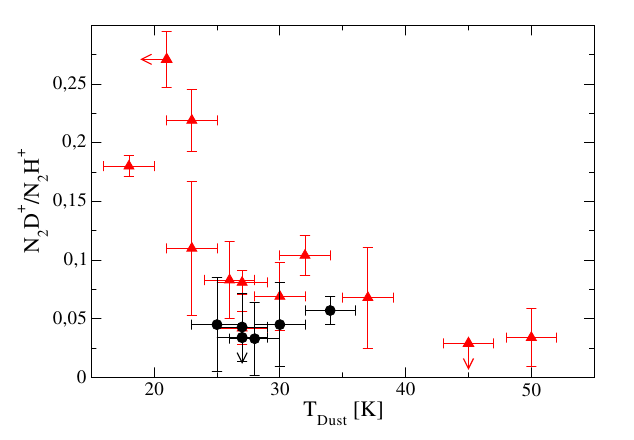}
\end{minipage}\hfil
\begin{minipage}{0.46\textwidth}
  \caption{\footnotesize Observations of \ntwodplus\ and \ntwohplus\ towards
    the low mass protostar in L1157 \citep{2013ApJ...765...18T}. The two
    species trace the flattened region of dense gas around the central
    protostar, with \ntwohplus\ (colour scale) peaking closer to the protostar
    than the \ntwodplus. The arrows indicate the direction of the outflow from
    the source.}
\label{fig_l1157}
\includegraphics[width=7.5cm]{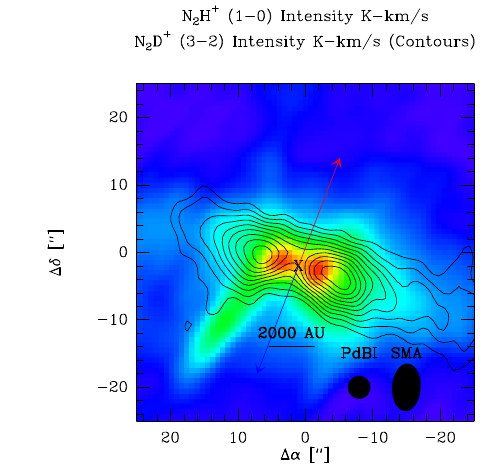}
\end{minipage}
\end{figure}

Deuterated species trace not only the quiescent gas contracting in prestellar
cores, but also the most quiescent regions around protostars; the gas which
has not yet been affected by feedback by the radiation or wind from a
protostar.  \citet{2009A&A...493...89E} have shown how the global \ntwodplus\
to \ntwohplus\ ratio decreases with increasing dust temperature providing an
age indicator for the youngest low mass protostars, the Class 0 sources
(Figure~\ref{fig_class0-evol}).  

Spatially resolving these deuterium rich circumstellar regions, as
\citet{2013ApJ...765...18T} have recently done for a small sample of low mass
protostars, will provide the opportunity to study pristine circumstellar gas.
Figure~\ref{fig_l1157} shows \citet{2013ApJ...765...18T} observations of the
protostar in L1157.  Here the \ntwodplus\ traces a cold, dense $\sim1000$
AU-sized flattened structure (a disk or toroid) around the central protostar
which drives a well collimated outflow in the perpendicular direction. The
increasing ratio of the \ntwohplus\ to \ntwodplus\ emission close to the star
is consistent with the heating of the material by the central source. However
the detailed analysis of of the \ntwodplus\ and its comparison with the
\ntwohplus\ are significantly hampered by the lack of \ntwodplus\ J=1-0 data,
which ALMA Band 2 can provide.

\begin{figure}[t]
\centering
\begin{minipage}{0.45\textwidth}
  \caption{\footnotesize Comparison of \ntwodplus\ and \ntwohplus\
    column density towards pre-stellar cores (small green squares;
    \citealt{2005ApJ...619..379C}), young high mass sources (blue
    triangles; \citealt{2011A&A...529L...7F}) and infrared dark clouds
    (large, red squares; \citealt{lackington13}). The dashed lines
    indicate lines of constant deuteration fraction. Note that the
    deuteration fractions plotted are lower limits on the true value
    as they observations have not been corrected for the smaller beam
    filling factor of the deuterated material than the \ntwohplus\
    emitting material in these single dish observations. }
\label{fig_dratios}
\includegraphics[width=7.5cm]{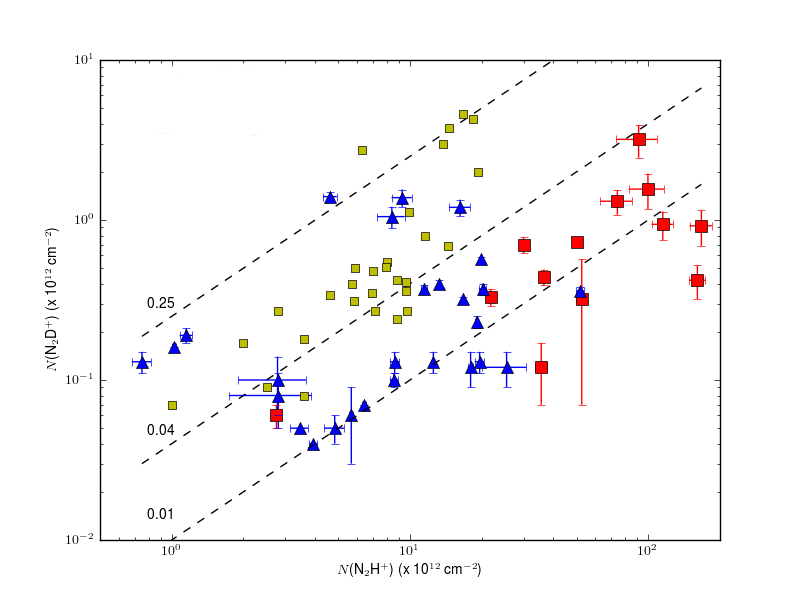}
\end{minipage}\hfil\hfil
\begin{minipage}{0.45\textwidth}
  \caption{\footnotesize The results of a model of deuterium fraction in the
    gas around a young massive star as a function of time since the start of
    heating by the star.  The thin lines show the ratio \ntwodplus\/\ntwohplus
    and while DNC/HNC is shown as thick lines.  The high initial deuteration
    fraction is quenced much more rapidly in \ntwodplus\/\ntwohplus than in
    DNC/HNC. The dotted and dashed curves correspond to different evolutionary
    ages of the gas when the heating starts \citep{2014MNRAS.440..448F}. }
\label{fig:dsp-ratio}
\includegraphics[width=7.5cm]{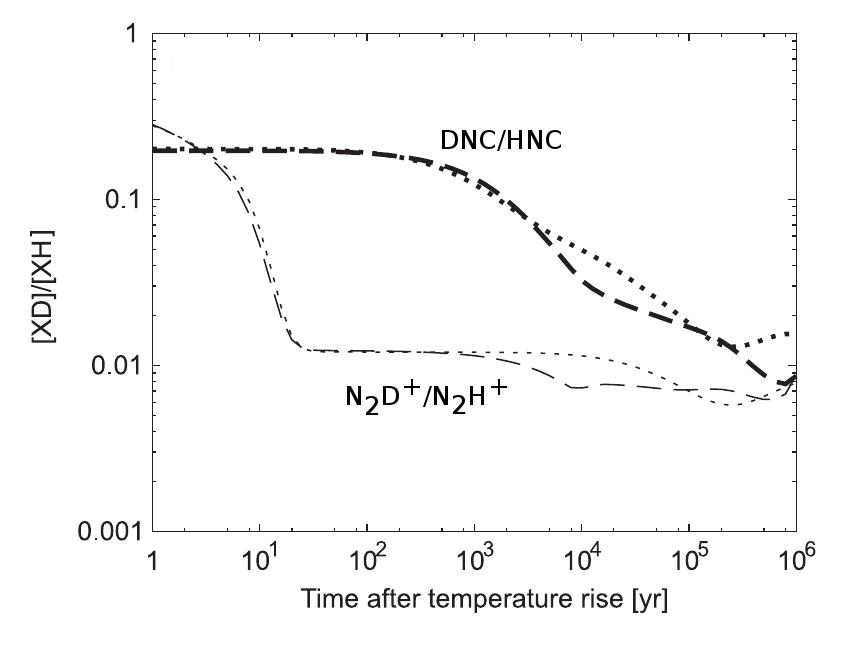}
\end{minipage}
\end{figure}

The conditions for high deuteration occur not only in
regions of low mass star formation but also in regions forming massive stars.
Observing a sample of objects spanning a likely range of evolutionary stages
in the formation of massive stars, from high mass starless core candidates to
ultra-compact \HII\ regions, \citet{2011A&A...529L...7F} detected deuteration
fractions in \ntwodplus\ as high as 25\%. These observations for the first
time identified deuteration fractions in massive star forming regions as high
as seen in low mass star forming regions.  These results demonstrate the
potential of deuterated species as a long searched for probe of the evolution
state and age of massive protostars.  Comparison of the deuteration fraction
for different species can also provide an estimate of the time since a young
massive has started to significantly heat its environment \citep[][;
  Figure~\ref{fig:dsp-ratio}]{2014MNRAS.440..448F}.  Once again, the isolated
hyperfine component of N$_2$D$^+$ will provide precious information on the
kinematics of the cold and dense gas on the verge of star
formation.

\subsection{Proto-planetary Disks: The CO Snowline}

Deuterated species are also important probes of the heavily shielded, and
therefore cold, dense mid-plane of proto-planetary disks; the regions within
which planets form. By tracing the decrease in gas phase CO abundance,
\ntwodplus\ and \ntwohplus\  can identify the location of the `CO snowline', an
important transition region for the properties of the dust. Outside the
snowline the frozen CO ice mantles enhance the `stickiness' of the dust,
promoting the growth of larger grains enroute to planet formation
\citep{2013A&A...552A.137R} and potentially impacting the composition of the
atmospheres of the planets which form \citep{2011ApJ...743L..16O}. The ice
also provides a substrate for chemical reactions which can alter the gas phase
chemistry, for example acting as a source of H$_2$CO
\citep{2013ApJ...765...34Q}.
 
Observations of DCO$^+$ together with other species, can map the ionization
fraction, a controlling parameter in the coupling of the gas and magnetic
field, as a function of radius and height in the disk
\citep{2011ApJ...743..152O}.  The presence of a quiescent `dead zone' within a
disk where the ionisation is too low for the magneto-rotational instability
(MRI) to operate would have important implications for the settling and
evolution of dust to form planetesimal and, once planets have formed, their
migration \citep{2007ApJ...654L.159C, 2009ApJ...691.1764M,
  2012ApJ...753L...8O}.

\begin{wrapfigure}{R}{8cm}
  \caption{\footnotesize Deuteration measured in various Solar System bodies
    \citep{2011Natur.478..218H}.}
\label{fig_solard}
\includegraphics[width=7.5cm]{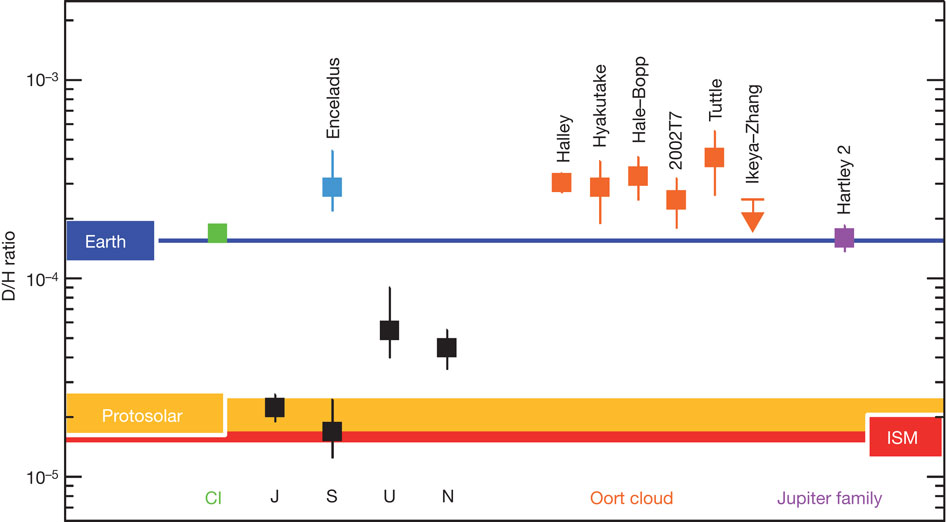}
\end{wrapfigure}

In our own solar system  a wide range of deuteration is observed
(Figure~\ref{fig_solard}) which has important implications for the origin and
evolution of volatiles, and especially water, on Earth: whether it is directly
accreted from the proto-solar nebula or delivered by asteroids and comets after
the Earth formed. Mapping the deuteration in proto-planetary disks can
potentially similarly constrain the transport of volatiles from the disk to
planets and how intra-planetary system variations are established.

\subsection{Deuterium in External Galaxies}
\label{sec:exgal-d}

\begin{figure}
  \caption{Simulated ALMA Band 2 spectrum of the J=1-0 transitions of
    deuterated species towards an external galaxy. The model assumes an
    excitation temperature of 10\,K, a line width of 100 km\,s$^{-1}$ and
    total H$_2$ column of $10^{23}$\,cm$^{-2}$. The abundance of the
    deuterated species is assumed to be $10^{-10}$, consistent with the models
    of \citet{2010ApJ...725..214B}. The rms noise level in the spectrum is
    1\,mK, consistent with a full ALMA observation of 35 minutes producing a
    5'' beam. 
}
\label{fig_exgal_dspec}
\centering
\includegraphics[width=0.8\textwidth]{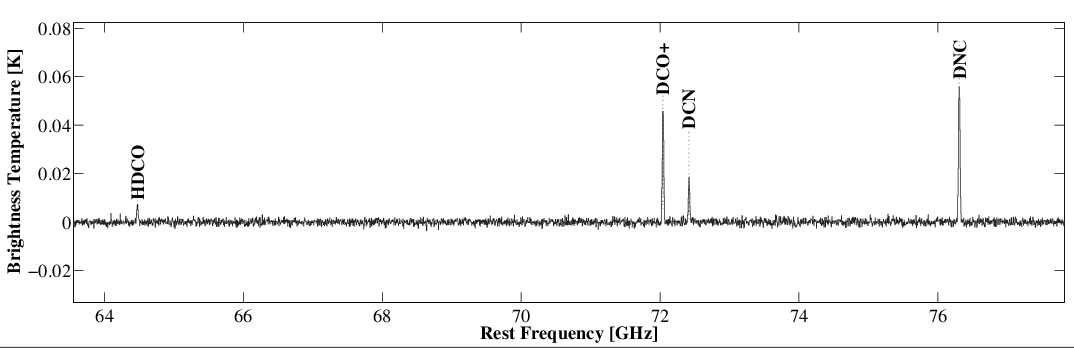}
\end{figure}

In the 40 years since the first molecular detection in the extragalactic
interstellar medium, the number of species identified in external galaxies is
now more than 50. Many of these species have been observed in nearby starburst
galaxies \citep{2006ApJS..164..450M,2011A&A...527A..36M,2011A&A...535A..84A, 2012JPhCS.372a2039A,2011EAS....52..299R}, due to their
brightness. Although the only current extragalactic detection of a deuterated
species, is a tentative detection of DCN by \citet{ 2006ApJS..164..450M} in
the nucleus of NGC253, ALMA Band 2 will revolutionise the study of these
species in external galaxies as the simulated spectrum in
Figure~\ref{fig_exgal_dspec} shows.

Astrochemical models \citep{2010ApJ...725..214B} have explored a large
parameter space of physical conditions covering different extragalactic
environments and provided a guide to observations of deuterated molecules
arising from the dense gas in external galaxies and the extreme environments
which they can probe. For example, the high spatial density of massive star
formation in mergers and starburst galaxies
\citep[e.g.][]{1993ApJ...413..542S,2009Natur.462..770V,2009Sci...326.1080A}
creates regions of extremely high cosmic ray energy density, up to about ten
thousand times that in the disk of the Milky Way Galaxy, potentially driving
the formation a top-heavy stellar initial mass function and bimodal star
formation \citep{2010ApJ...720..226P, 2011MNRAS.414.1705P}.

Although currently qualitative in nature, the astrochemistry models
provide some important conclusions: (i) HDO and DCN are abundant,
regardless of the extragalactic environment; and (ii) DCO$^+$ is a
tracer of enhanced cosmic-ray irradiated molecular gas which together
with the other ions accessible in the ALMA Band 2+3 (HCO$^+$, HOC$^+$,
\ntwohplus\ and their deuterated isotopologues), probes the ionization
of the gas.

\subsection{A Second Route to Deuteration?}

One surprising result is the high levels of deuteration seen in DCN in
some regions of warm gas such as in the inner regions of circumstellar
disks \citep{2012ApJ...749..162O} and in the Orion Bar
\citep{2009A&A...508..737P}. Two possible explanations have been
advanced for these observations. First, the gas phase DCN is
transitory, resulting from the current evaporation of DCN formed in
the ice mantles during the earlier evolution of the gas and dust
through a cold, dense phase
\citep{1983A&A...119..177T}. Alternatively, a deuteration route
involving CH$_2$D$^+$, which is effective for temperatures up to
$70$\,K, could directly produce high levels of deuteration in the warm
gas \citep{2009A&A...508..737P,2012ApJ...749..162O}. Distinguishing
between these two possible explanations in a given region requires
observations of CH$_2$D$^+$ and ALMA Band 2 is essential in doing this
as it contains the J=1-0 transition of CH$_2$D$^+$
(Table~\ref{tab_dspecies}).

\subsection{ALMA Band 2: The Deuterium Band}

\begin{figure}[bht]
\centering
\includegraphics[width=0.5\textwidth]{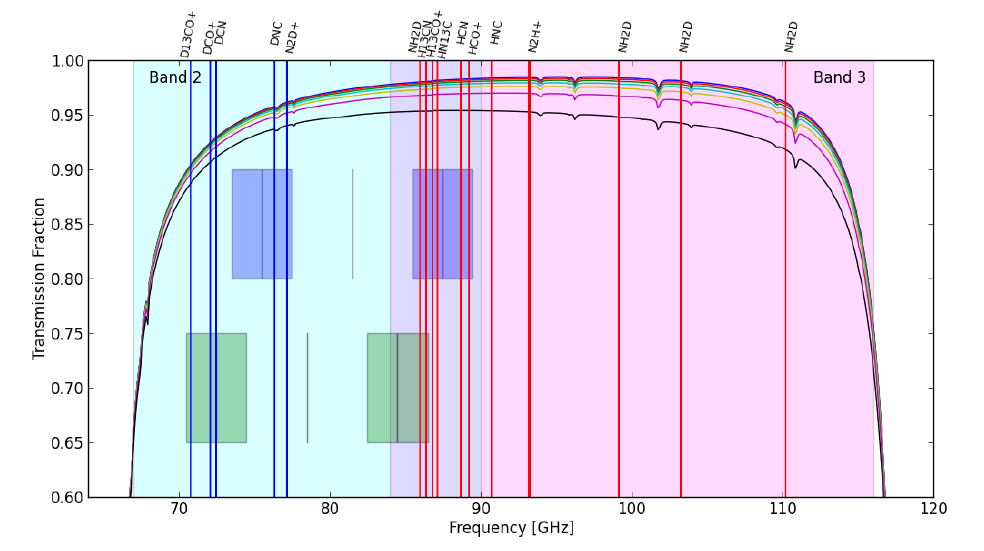}\includegraphics[width=0.5\textwidth]{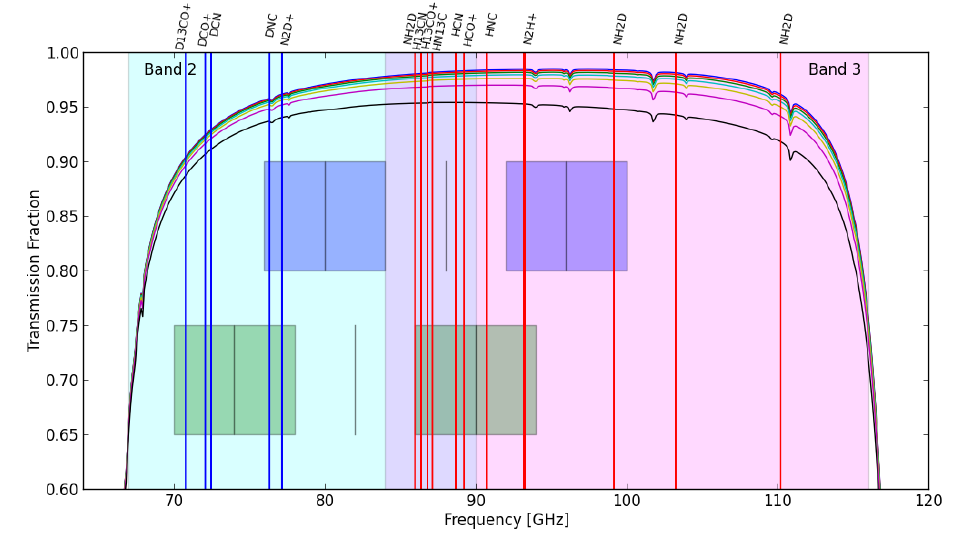}
\caption{\footnotesize Possible frequency setups to observe deuterated
  molecules and closely related species with a Band 2+3 receiver
  system. \emph{Left:} Two setups with the current 4x2GHz basebands. Setup 1,
  shown in green, covers the J=1-0 transitions of D$^{13}$CO$^+$, DCO$^+$, DCN,
  \nhtwod and H$^{13}$CN. Setup 2,
  shown in blue, covers the J=1-0 transitions of DNC, \ntwodplus, \nhtwod,
  H$^{13}$CN, H$^{13}$CO$^+$, HN$^{13}$C, HCN and HCO$^+$. \emph{Right:} Two
  setups with a future 4x4GHz basebands system. Setup 1,
  shown in green, covers the J=1-0 transitions of D$^{13}$CO$^+$, DCO$^+$, DCN,
  DNC, \ntwodplus, \nhtwod, H$^{13}$CN,  H$^{13}$CO$^+$, HN$^{13}$C, HCO$^+$,
  HCN and \ntwohplus. Setup 2,
  shown in blue, covers the J=1-0 transitions of DNC, \ntwodplus,
  \ntwohplus and \nhtwod.}
\label{fig:freqsetups}
\end{figure}

The enhancement of deuterated species in cold regions directly points
to the importance of the low J transitions, and in particular the
J=1-0 transition, in reliably establishing the excitation and column
density of deuterated species.  Many of the simple deuterated species
including \ntwodplus, DCO$^+$, DCN, DNC, DOC$^+$ (as well as their
$^{13}$C and $^{15}$N containing isotopologues) have their ground
state transitions in ALMA Band 2 (Table~\ref{tab_dspecies}). Among the
ALMA frequency bands, Band 2 is unique in its ability to observe many
of these important deuterated species simultaneously. In a single
frequency (Figure~\ref{fig:freqsetups}, left) setting DCO$^+$ and DCN
and their $^{13}$C and $^{15}$N isotopologues can be observed together
with N$_2$D (deutero-ammonia). While a second setting
(Figure~\ref{fig:freqsetups}, right) can observe DOC$^+$, DNC,
\ntwodplus\ and N$_2$D together with the $^{13}$C variants of HCN,
HCO$^+$ and HNC. A future upgrade to a 16 GHz bandwidth will
allow simultaneous observation of all these deuterated species plus
\ntwohplus in a single setting, as well as HCN, HNC HCO$^+$ and their
$^{13}$C and $^{15}$N isotopologues. No other ALMA frequency band can
match Band 2 in this ability to study deuterated molecules, both in
terms of the number of species accessible and the importance of the
transitions. \todo[noline]{ Model to justify importance of
  J=1-0 lines in constraining column densities and excitation. Also
  describe efficiency by pointing out difficulty in observing lines in
other bands.}

\section{Level 2 Science Projects}

\subsection{The nature of the earliest ionized gas during
  massive star formation}
\todo[noline]{Add figure.}  Ultracompact \HII\ regions (\UCHIIs) have been
known for decades to be the clearest signpost of the formation of truly
massive stars (O and early B, M$>10$M$_\odot$), because only these stars can
produce a significant numbers of UV photons ($N>10^{43}$~s$^{-1}$).  Typical
\UCHIIs\ discovered in early surveys \citep[e.g.][]{1989ApJ...340..265W} have
emission measures $\mathcal{E}=ln_e^2 \sim 10^6 \textrm{ to }
10^7$~pc~cm$^{-6}$ (where $l$ and $n_e$ are the size and electron number
density respectively) and can well be accounted for by homogeneous ionized gas
that becomes optically thin (marked by a spectral index change from $\sim2$ in
the optically thick regime to $\sim-0.1$) at a frequency of 1 to 5~GHz.
However, a population of smaller (sizes of hundreds to few thousands of AU),
denser \HII\ regions has been identified, usually called hypercompact \HII\
regions (HCHIIs).  The turnover frequency of a homogeneous \HII\ region scales
as $\nu_{\rm turnover} \propto \mathcal{E}^{0.5}$
\citep[see][]{2005IAUS..227..111K}, so the turnover frequency of these \HII\
regions with $\mathcal{E}\sim 10^9 \textrm{ to } 10^{11}$~pc~cm$^{-6}$ falls
in the range from 20 to 150 GHz, the central part of which would be covered by
ALMA Band 2 (65 to 90 GHz).  These \HCHIIs\ are believed to signpost the
appearance of photoionization in the formation of O-type and early B massive
stars, and therefore are key to understanding the late stages of formation and
end of accretion.  There is evidence that \HCHII\ regions are not simply
homogeneous spheres expanding hydrodynamically, but that they have density
gradients \citep[see e.g. Fig. 6 of][]{2008ApJ...672..423K} and supersonic
motions (broad recombination lines with FWHM$>20$ km~s$^{-1}$) due to ionized
outflow or even inflow \citep[e.g.][]{2006ApJ...637..850K}.  ALMA with Band 2
capabilities would offer the unique opportunity of characterizing this early
ionized gas both in free-free continuum (SEDs) and in hydrogen recombination
lines. The latter are a unique probe of the dynamics of the ionized gas.
Furthermore, mm recombination lines cover a unique niche compared to cm and
sub-mm RLs: the cm lines are faint and are pressure broadened, whereas the
sub-mm RLs can be contaminated with molecular lines and are more prone to
non-LTE effects \citep[e.g.][]{2012MNRAS.425.2352P,2012A&A...547L...3G}.  ALMA
Band 2 is able to observe the alpha RLs from H45$\alpha$ (69.830~GHz) to
H42$\alpha$ (85.688~GHz) while a Band 2+3 system will reach up to H37$\alpha$
(106.737~GHz) and H38$\alpha$ (115.274~GHz). A Band 2+3 system with a 16 GHz
bandwidth would allow the simultaneous observation of three recombination
lines (H45$\alpha$, 46$\alpha$ and 41$\alpha$) which, because of the
increasing frequency separation of the higher $\alpha$ lines, is not possible
in any other ALMA Band. At a typical distance of 5 kpc, the maximum ALMA
angular resolution in this band of 0.05" is equivalent to $\sim250$AU,
sufficient to map the continuum and dynamics of these sources.
\todo{Demonstrate feasibility.}

\subsection{From Grains to Planets: The evolution of dust from the
  interstellar medium to planet formation}

The far-IR to mm wavelength spectral energy distribution (SED) of the
interstellar medium is dominated by thermal emission from dust grains with the
shape of the SED sensitive to the properties of the dust grains. In
particular, for a given dust grain composition and shape, the slope of the
emission flux density as a function of wavelength traces the dust grain size
distribution, becoming shallower as the grains become
larger \citep{2003ARA&A..41..241D}. By measuring the variation in this spectral
slope across the (sub)mm wavelength window from the pristine interstellar
medium to protoplanetary disks it is possible to trace the evolution of the
interstellar dust as it grows to form planets
\citep{2006ApJ...636.1114D,2007prpl.conf..767N}.

\begin{wrapfigure}{r}{8cm}
\vspace*{-0.5cm}
\includegraphics[width=.45\columnwidth]{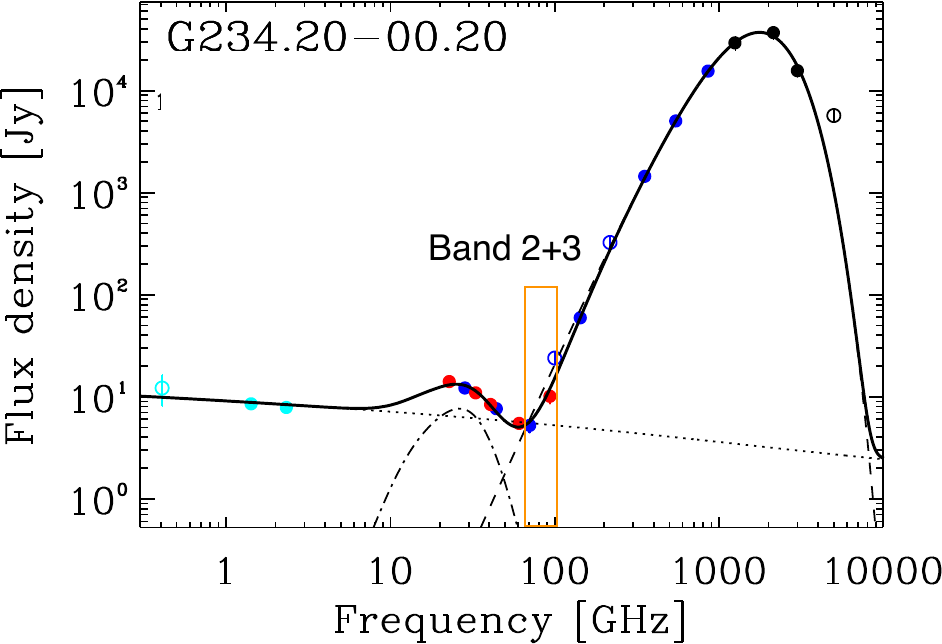}
\caption{\footnotesize The spectral energy distribution of a region showing the emission from spinning dust peaking at $\sim20$\,GHz \citep{2014A&A...565A.103P}. The orange box shows the region covered by Band 2+3 which spans the minimum in emission between the thermal dust emission at higher frequencies and spinning dust emission at lower frequencies.  }
\label{fig:dustsed}
\end{wrapfigure}

The fact that observations are most sensitive to the emission from
grains with sizes similar to the observing wavelength, means there is
a considerable advantage in measuring grain growth by observing at
longer wavelengths. However, around cm wavelengths the contribution to
the SED from free-free emission and spinning dust begins to dominate
over the thermal dust emission. It is important to accurately model
and remove this emission.  In low mass star forming regions and
protoplanetary disks, the ALMA Band 2 window is the longest wavelength
still dominated by thermal dust emission (Figure~\ref{fig:dustsed}). Combined with ALMA Band 1 as
well as longer wavelength VLA/SKA observations ALMA Band 2 will be a
uniquely powerful probe of grain physics and growth in protoplanetary
disks. We stress that full coverage of the spectral energy
distribution from the sub-mm through cm wavelengths, and especially
across the 100-30~GHz frequency range\todo{Justify. Figure?}, is
essential to disentangle the various contributions of the emission.

 In order to sample the SED across the
full sub-mm to cm wavelength window, previous measurements of grain
growth have needed to rely on combining flux measurements from
observations at different telescopes. The absolute flux calibration
has been a major factor limiting the accuracy with which it is
possible to derive the spectral slope, and hence dust grain size
distribution. One immediate advantage of ALMA in this area of science
is the much greater absolute flux calibration accuracy over existing
facilities. An additional advantage of going to long wavelengths, is
that for a fixed bandwidth (as is the case for ALMA continuum
observations) the fractional bandwidth in a single wavelength setup
increases. In other words, for a source detected at a given signal to
noise, it becomes easier to detect an intrinsic spectral index in the
source across the instantaneously-observed bandpass. This has the
major advantage that the uncertainty in the measured spectral slope of
the source is limited by the signal to noise per channel and bandpass
calibration and \emph{not} the absolute flux calibration. If the
instantaneous bandwidth is increased from 8 to 16\,GHz, as proposed in
one part of the ALMA development plan, or if the Band 2 receivers
extend to cover the Band 3 wavelength window, the advantages outlined
above improve dramatically. This accuracy is expected to allow precise
measurements of grain size variations within disks, which are a
powerful probe of the grain growth physics and its evolution towards
planet formation \citep{2012A&A...538A.114P,2013Sci...340.1199V}.

Typical star-forming cores in nearby star formation regions have
angular sizes of $\sim30$". The large primary beam (70--90'') at Band 2
makes it possible to observe these in a single pointing without
worrying about primary beam attenuation or the reduced surface
brightness sensitivity penalty arising when mosaiced observations are
required.

The expected very good phase stability at Band 2 wavelengths means it
should be possible to image with the most extended antenna
configurations, and so achieve angular resolutions
$<0.1$". \todo{Demonstrate with simulations.} This
is particularly important for studying dust properties in
protoplanetary disks. Grain growth is predicted to be non-uniform
across the disk, varying both generally with radius in the disk, but
also due to stochastic physical processes
\citep{2012ApJ...760L..17P,2013Sci...340.1199V}. Being able to
spatially resolve the dust grain size distribution across the disk
would help address a key problem in the understanding of planet
formation -- the `metre barrier'. Theoretically we understand how
small grains can grow to metre sizes in a disk by
coagulation. However, current turbulent disk models struggle to make
particles much bigger than this as these large metre-size boulders get
destroyed by collisions. One potential mechanism to overcome the
barrier is that gas and dust get pressure-confined in small pockets
within the disk, allowing the dust grains to interact constructively
rather than destructively, and thereby grow to sizes larger than the
theoretical barrier. Once they do so they are no longer destroyed by
collisions and can grow to form planetesimals and eventually
planets. The upper-limit to the `barrier size' varies inversely
linearly proportionally with the radius from the central star. At
1\,AU from the star, the barrier occurs for grains of metre
size. While at 100\,AU the barrier appears to limit grains to sizes
below millimetre sizes. One can use ALMA Band 2 observations to
identify the regions in the outer disk where grains have grown from mm
to cm sizes. Once identified, studying these regions to understand how
the grains overcome the barrier will be a major step forward in
understanding planet formation, including the formation of earth-like
planets close to the habitable zone if the same physical processes
operate at 1\,AU as at 100\,AU.

\subsection{Isotopic variations: Probing the mixing in stars}

The fundamental role played by asymptotic giant branch (AGB) stars in
the enrichment of the interstellar medium (ISM) and the chemical
evolution of galaxies is well-established
\citep{1999ARA&A..37..239B}. Almost half of the heavy elements
returned to the ISM originates in evolved stars with initial masses
between 1 and 8\,M$_\odot$ \citep{1992A&A...264..105M}. The chemical
composition of the material returned to the ISM during the AGB, is
determined by the nucleosynthesis in the stellar core and the
subsequent convective envelope mixing. Isotopic ratios are the best
diagnostic tracers for the efficiency of the nucleosynthesis and
mixing processes, as they are very sensitive to the precise conditions
in the nuclear burning regions, and observations of the different
molecular isotopologues are therefore the key to understanding the
stellar origin of elements. Despite their obvious importance, the
observational constraints are scarce, mainly due to the weakness of
the various emission lines, and very little recent progress has been
made. Scattered observations exist of small samples and essentially a
statistically relevant data set is only availble for the carbon
isotopes \citep{1986ApJS...62..373L,2000A&A...359..586S} and to some
extent, for the important oxygen isotopes \citep{1987ApJ...316..294H},
necessary to constrain the mixing processes. For the silicon isotopic
ratios, essential to understand the origin of presolar SiC grains and
the enrichment of our own Solar system
\citep[e.g.][]{2013ApJ...768L..19L}, the situation is even worse, with
observations available only for a handful of stars. With its
unprecedented sensitivity, ALMA can play a crucial role in this field,
fundamental to stellar evolution and the formation of elements, by
constraining the isotopic ratios in statistically significant samples
of AGB stars with a limited amount of observing time.
\begin{table}
\centering
\caption{Transitions of silicon containing isotopologues in ALMA Band 2.\label{tab:si}}
\begin{tabular}{ccccccc}
\toprule
Species & Transition & Frequency (GHz)& & Species & Transition & Frequency
(GHz)\\
\midrule
Si$^{13}$CC & 3(1,3)-2(1,2) & 65.036 &           &   Si$^{13}$CC & 3(1,2)-2(1,1) & 73.102 \\
$^{30}$Si$^{34}$S v=0 & J=J=4-3 & 68.052 & 	 &   SiC$_2$ v=0 & 21(4,17)-21(4,18) & 73.178 \\	
$^{30}$SiC$_2$ & 3(0,3)-2(0,2) & 68.333 & 	 &   Si$^{13}$CC & 7(1,6)-7(1,7) & 74.384 \\ 
Si$^{13}$CC & 3(0,3)-2(0,2) & 68.610 & 		 &   Si$^{18}$O v=0 & J=2-1 & 80.705 \\ 	    
$^{30}$SiC$_2$ & 3(2,2)-2(2,1) & 68.777 & 	 &   $^{30}$SiO v=2 & J=2-1 & 83.583 \\ 	    	
Si$^{13}$CC & 3(2,2)-2(2,1) & 69.129 & 		 &   $^{30}$SiO v=1 & J=2-1 & 84.164 \\      
$^{30}$SiC$_2$ & 3(2,1)-2(2,0) & 69.255 & 	 &   SiO v=4 & J=2-1 & 84.436 \\	      	    	
$^{29}$SiC$_2$ & 3(0,3)-2(0,2) & 69.264 & 	 &   $^{29}$SiO v=2 & J=2-1 & 84.575 \\      	
Si$^{13}$CC & 3(2,1)-2(2,0) & 69.682 & 		 &   $^{30}$SiO v=0 & J=2-1 & 84.746 \\        	
$^{29}$SiC$_2$ & 3(2,2)-2(2,1) & 69.735 & 	 &   SiO v=3 & J=2-1 & 85.038 \\ 	
SiC$_2$ v=0 & 11(2,9)-11(2,10) & 69.910 & 	 &   $^{30}$Si$^{34}$S v=0 & J=5-4 & 85.065 \\	
$^{30}$SiS v=0 & J=4-3 & 70.041 & 		 &   $^{29}$SiO v=1 & J=2-1 & 85.167 \\        	
$^{29}$SiC$_2$ & 3(2,1)-2(2,0) & 70.242 & 	 &   SiO v=2 & J=2-1 & 85.640 \\ 	    	
SiC$_2$ v=0 & 3(0,3)-2(0,2) & 70.260 & 		 &   $^{29}$SiO v=0 & J=2-1 & 85.759 \\        	
Si$^{34}$S v=0 & J=4-3 & 70.629 & 		 &   SiO v=1 & J=2-1 & 86.243 \\ 	    	
SiC$_2$ v=0 & 3(2,2)-2(2,1) & 70.763 & 		 &   Si$^{13}$CC & 4(1,4)-3(1,3) & 86.563 \\   	
$^{29}$SiS v=0 & J=4-3 & 71.284 & 		 &   SiO v=0 & J=2-1 & 86.847 \\ 	    
SiC$_2$ v=0 & 3(21)-2(20) & 71.302 & 		 &   $^{30}$SiS v=0 & J=5-4 & 87.551 \\     	
SiS v=3 & J=4-3 & 71.558 & 			 &   Si$^{34}$S v=0 & J=5-4 & 88.286 \\    	
Si$^{33}$S v=0 & J=4-3 & 71.595 & 		 &   $^{29}$SiS v=0 & J=5-4 & 89.104 \\     	
SiS v=2 & J=4-3 & 71.911 & 			 &   SiS v=3 & J=5-4 & 89.446 \\ 	      	
SiS v=1 & J=4-3 & 72.265 & 			 &   Si$^{33}$S v=0 & J=5-4 & 89.489 \\     	
SiS v=0 & J=4-3 & 72.618 & 			 &   SiS v=2 & J=5-4 & 89.888\\
\bottomrule
\end{tabular}
\end{table}

%

With a combined Band 2+3 receiver, simultaneous observations of several
different important isotopologues of the most abundant elements can be
obtained in one observational set-up. Simultaneous observations of lower
frequency transitions are crucial for reliable abundance estimates, since they
probe the outflowing cool gas on the largest scales. In particular, some of
the lower transitions of the oxygen, silicon and sulfur isotopologues of SiO
and SiS are only available within the frequency range of Band 2
(Table~\ref{tab:si}).

The circumstellar envelopes of mass-losing AGB stars within 500 pc
(constituting a sample of a few hundred stars) range from a few to several
tens of arcsecs on the largest scales. The larger primary beam at the lower
frequencies is well matched to the size of the envelopes of the most
extended, allowing their imaging without concerns about the primary
beam attenuation or the need for mosaic observations. 

\subsection{Sunyaev-Zel'dovich Effect Observations of Galaxy Clusters\label{sec:sz}}

Galaxy clusters are among the largest gravitationally bound structures to form
by the current epoch, and serve both as probes of cosmology and astrophysical
laboratories for tests of plasma physics, shocks, the particle nature of dark 
matter, AGN lifecycles and feedback, and the impact of environment on galaxy
evolution, to name a few examples.
One exciting probe of the intracluster medium (ICM) -- the hot ($\sim 10^7-10^8$~K)
X-ray emitting gas that comprises $\sim 15\%$ of a cluster's total mass -- is the thermal 
Sunyaev-Zel'dovich (SZ) effect (\cite{sunyaev1972}; 
for reviews, see \cite{2002ARA&A..40..643C,2014PTEP.2014fB111K}).  Due to the inverse Compton scattering of 
photons from the Cosmic Microwave Background (CMB), the SZ effect is 
proportional to the Compton $y$ parameter, which scales as the line of
sight integral of electron pressure ($y \propto \int P_e d\ell$).
The SZ effect uniquely has redshift-independent surface brightness and manifests itself
as a decrement in intensity (or brightness temperature) at frequencies below the peak in the 
CMB spectrum ($\lesssim 218$~GHz).  By contrast, X-ray emission suffers from
cosmological dimming and scales as the line of sight integral of ICM density squared,
meaning X-ray surface brightness declines rapidly with both cluster radius and redshift.

\begin{figure}[htb!]
\begin{center}
\includegraphics[width=0.8\columnwidth]{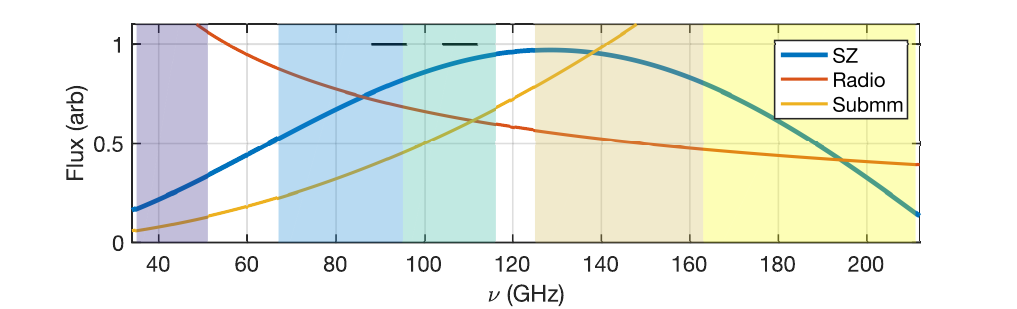}
\includegraphics[width=0.8\columnwidth]{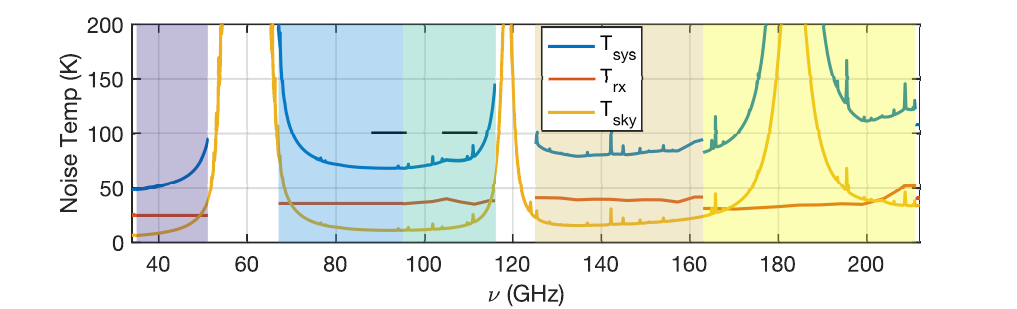}
\includegraphics[width=0.8\columnwidth]{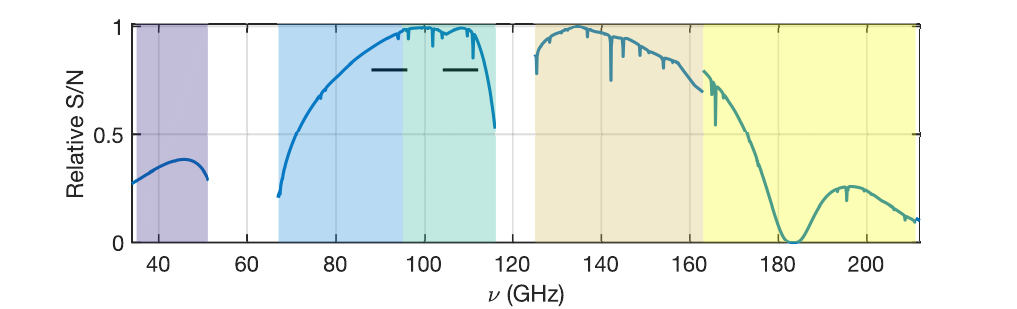}
\end{center}
\caption{\footnotesize
For all plots, ALMA Bands 1--5 are indicated as colored regions in the background.  Note that we chose 95 GHz as the upper end of Band 2, as this corresponds to a proposed Band 2+ design for ALMA and is the portion of the spectrum with tighter specifications for Band 2+3.
{\it Upper panel:} The blue curve shows the magnitude of the SZ flux density as a function of frequency for a feature of fixed angular extent, with amplitude scaled arbitrarily. 
Contributions from various emission mechanisms, which may contaminate SZ signals, are also plotted \citep{2013ApJS..208...20B}. 
The red curve represents the spectrum of a typical unresolved radio source with
a power law spectrum of $\nu\approx-0.7$, while the orange-yellow  curve represents
a typical dusty submillimeter galaxy in or behind the cluster ($\nu\approx2$), also arbitrarily scaled.
Such contamination is expected to reach a minimum in Band 2+3.
{\it Middle panel:} Receiver and sky noise temperature as a function of frequency for
the bands spanning the SZ decrement ($\nu \lesssim 218$~GHz, ALMA Bands 1--5).
The values of precipitable water vapour (PWV) content in the atmospheric model
(using Scott Paine's AM code; \cite{paine_scott_2017_438726}) is 2.748~mm (ALMA 6$^{th}$ octile weather).
{\it Lower panel:} Relative signal to noise on the SZ decrement after accounting for the sky and receiver noise temperatures, conservatively assuming  a receiver noise temperature of 36 K in Band 2+.  An optimally-sensitive Band 2+3 receiver tuning with two 8 GHz sidebands (16 GHz total bandwidth) spanning 24 GHz on sky is shown in black. 
Such a configuration is likely when accounting for ALMA upgrades under study that will digitize a 4-12 GHz IF band. 
This tuning is not possible with current Band 3 or the proposed Band 2+.}
\label{fig:SZnew}
\end{figure}

The SZ spectrum is a broadband continuum, resolvable over the same broad range of angular scales subtended by a galaxy cluster.
Figure \ref{fig:SZnew} (top) shows the relative magnitude of the SZ intensity in ALMA Bands 1--5 
(i.e.\ $\nu<218$~GHz, where the SZ effect is a decrement), 
compared to typical dusty submm and radio synchrotron sources that can contaminate SZ measurements.
Because the angular diameter distance $d_A(z)$ is a slow function of redshift at 
intermediate and higher redshifts ($z\gtrsim0.2$), galaxy clusters at these
distances typically subtend several arcminutes.  The exception to this
is the highest redshift clusters.
Growing from the largest fluctuations in the primordial matter density power spectrum,
these systems serve as powerful probes of cosmology and cluster evolution, and
are lower mass than their nearby counterparts.
Thanks to deep SZ, X-ray, and near-IR surveys, such systems are now being discovered 
at $z\sim2$, which is near the turnover in $d_A(z)$. 
Here a cluster of a given physical size will subtend the 
smallest angular size, and the bulk SZ signal is often contained within 1-2'.
Additionally, submillimeter surveys have discovered protoclusters (lower mass 
cluster precursors) as associations of dusty, star-forming galaxies. 
Bands 2 and 2+3 will serve as a powerful and sensitive tool for probing the SZ effect
from such systems as they begin to virialize (Figure \ref{fig:SZnew}, lower panel).
ALMA's sensitivity in Band 2 may also allow detection of the SZ effect from individual galaxies. 
Massardi et al.\ 2008 predicted tens to hundreds of galaxies could be found in a 1 sq.\ degree 
survey with Band 2 through their SZ signatures, which would subtend $\sim10''$ scales.
By stacking the  {\it Planck}  SZ signals from the locally brightest galaxies, 
recent work by Greco et al.\ 2015 has shown that galaxies with masses $>10^{11}~\rm M_\odot$ 
have SZ fluxes on the order of several tens to hundreds of $\mu$Jy in Band 2,
readily detectable in short ALMA observations.

Since ALMA Band 1 will have the largest field of view ($\sim 3'$), 
and therefore will recover the largest angular extents, 
the case has been made that it will be the best ALMA band for detection of 
galaxy clusters through the SZ effect (see e.g.\ the ALMA Band 1 science case).
This argument strongly favors brightness temperature, which benefits
from the larger beam size useful for detection of an extended source.
However, for astrophysical studies we are often interested in features of a 
given angular scale or in a range of scales, and for that one must consider
the flux density on these scales.
Figure \ref{fig:SZnew} (lower panel) shows that for studies of SZ 
features of a given angular size, a broad tuning that spans Band 2+3 
will offer the highest sensitivity. The tuning setup shown considers likely 
upgrades to the ALMA digitizers and correlator, which will bring 16-GHz of
total bandwidth in two 8-GHz sidebands separated by 8 GHz (i.e.\ by
digitizing IF bands spanning 4-12 GHz).

A few of the cluster astrophysical studies that will benefit by the 
construction Band 2+3 are:

\begin{itemize}

\item {\bf Studies of Shocks:}
Galaxy clusters assemble hierarchically by accreting filamentary gas and 
merging subclusters that drive shocks through the ICM.  
Of the three main thermodynamic parameters -- temperature, density, and pressure 
-- pressure is the most sensitive to a shock's Mach number. 
As demonstrated in MUSTANG studies exploiting the 9'' resolution of the 100-meter Green
Bank Telescope (GBT) at 3.3~mm (e.g.\ \cite{2010ApJ...716..739M,2011ApJ...734...10K} and more 
recently in ALMA Band 3 observations (e.g.\ \cite{2016PASJ...68...88K,2016ApJ...829L..23B}), the SZ is a natural tool 
for detecting and characterizing the pressure discontinuity at a shock, particularly
in high-$z$ clusters where cosmological dimming of X-ray surface brightness makes 
X-ray observations challenging and expensive. The differing line of sight dependence
of the thermal SZ effect also offers a way to test X-ray modeling of shocks.

\item {\bf Studies of Cold Fronts:}
In many merging clusters, contact density discontinuities with no clear, associated 
pressure jumps have been observed.  
This is expected both inside the shock front where the density
gradient steepens and in subsonic motions such as minor mergers that induce sloshing.
In such features, called ``cold fronts'', the jump in density is accompanied by a 
corresponding drop in temperature, keeping the thermal pressure roughly constant.
If SZ observations were to reveal a discontinuity at a cold front, this would indicate some
of the pressure balance across the front is due to non-thermal pressure, such as
that due to turbulence and magnetic fields.

\item {\bf Studies of AGN-driven Radio Bubbles / X-ray Cavities:} 
Feedback from AGN offsetting cooling flows is thought to occur primarily mechanically,
through shocks driven by AGN jet-inflated bubbles.  The SZ can both probe the shocks
and, uniquely, determine the plasma composition in these radio bubbles (see \cite{2005A&A...430..799P}).  
Observations of the SZ signature from X-ray cavities can
provide valuable insight into the nature of AGN feedback in galaxy clusters.

\item {\bf Studies of ICM Turbulence through Pressure Fluctuations:} 
Turbulent motions in the intracluster medium have long been an open question
for cluster astrophysics. These motions can contribute to the non-thermal 
pressure support not accounted for in  X-ray mass estimates that assume thermal
hydrostatic equilibrium (HSE).  
Turbulent motions in the ICM are thought to bias mass estimates 
by $15-20\%$ at radii typically used in cosmological studies.\footnote{This radius is 
typically $R_{500}$, the radius within which the average density of the cluster exceeds 
the critical density of the Universe at the cluster redshift by 500$\times$.} 
The nature of turbulent motions can also be used to place constraints on ICM viscosity
and the scales of injection and dissipation, which are important parameters in 
hydrodynamic simulations that include cooling, mixing, and cluster magnetic fields.

Cluster astrophysicists have long sought for an X-ray microcalorimeter observatory
to measure ICM motion (through X-ray line spectroscopy) and spectroscopic measurements of
temperature fluctuations.  
However, with the loss of the {\it Hitomi} (Astro-H) X-ray satellite, the next opportunity 
for such an instrument may come no earlier than 2028 with the planned launch date for {\it Athena}.
Measuring pressure fluctuations through the SZ power spectrum of galaxy clusters will probe 
cluster outskirts, which matter for reliable mass estimation, in the much nearer term, advancing
the state of the art for ICM physics now.

\item {\bf Joint SZ/X-ray Surface Brightness Constraints on Cluster Temperature:}
The eROSITA X-ray satellite will soon launch, and will perform the first all-sky X-ray survey
in nearly 3 decades.  It is expected to detect $>7\times10^4$ clusters and groups of galaxies out to  
$z\sim1.3$ (\cite{2012MNRAS.422...44P,2015IAUGA..2257162P}.  
With a detection limit of $\gtrsim$50 photons, few of these detections will accumulate enough 
counts for X-ray spectroscopy.  SZ flux measurements can provide confirmation, mass estimates
through scaling relations, and a method for determining a cluster's radial temperature profile.
The combination of SZ and X-ray flux allows one to jointly probe cluster temperature 
(see e.g.\ \cite{2009ApJ...694.1034M}) without relying on spectroscopy.  
Further, this combination can also be used 
to estimate roughly the cluster redshift (see \cite{2015MNRAS.450.1984C}).

\item {\bf High redshift clusters and protoclusters}
Detection of the SZ effect from forming high-$z$ clusters and protoclusters
with compact angular size, well-matched to the beam size in Band 2+3.

\end{itemize}

Recent studies (e.g. \cite{2013ApJ...764..152S,2014MNRAS.445..460G,2014A&A...565A.103P}) 
have shown that Band 2+3 is near the minimum in contamination from radio and submillimeter sources, while the broad
coverage of Bands 2+3 will offer better spectral leverage for separating contamination from the SZ
signal.  For good sampling of a broad range of spatial scales in a cluster, which is required
both for spatial dynamic range in imaging and for studies of the pressure power spectrum
in clusters, the broad fractional bandwidth ($>50\%$) offered by Bands 2+ and 2+3 will offer exceptionally good {\it uv}-coverage.
Such high-fidelity imaging at high resolution ($\sim$5'' resolution, compared to the typical arcminute
resolution of an SZ survey instrument) is also important for probing the sources of scatter
in SZ--mass scaling relations used for cosmology.  

\subsection{VLBI}

The millimetre very long baseline interferometry (VLBI) science enabled by a
beamformer for ALMA has recently been discussed by \citet{2013arXiv1309.3519F}.
Below we discuss the range of potential VLBI science applications for Band 2
or Band 2+3 with ALMA in conjunction with other millimetre
telescopes. Although few other millimetre telescopes currently operate over
the full frequency range of ALMA Band 2/2+3, the availability of this band and
the resolution, baseline coverage and sensitivity which ALMA will be able to
provide for VLBI will likely stimulate adoption of this frequency range at
other telescopes. At IRAM receivers which extend well in to the Band 2
frequency range are already in development.

\subsubsection{Masers}

The intense and spatially compact emission from masers allows the study of
regions at resolutions not possible in other, thermally excited, molecular
lines. Several probable maser transitions of methanol \citep[66.95, 68.31 GHz;
Class II masers;][]{2005MNRAS.360..533C} and formaldehyde (72.84 GHz), fall in
ALMA Band 2.  (A further four Class II and three Class I methanol masers are
known in Band 3 \citep{2004A&A...428.1019M}.)  Some of these lines have been
detected (from Galactic star forming regions) with single dishes but knowledge
of this spectral region is limited due to the current lack of facilities.
Models of methanol by \citet{2005MNRAS.360..533C} are successful in explaining
the location of masers which have been imaged at high resolution and the
detection statistics and line profiles can be used to develop diagnostics for
maser locations and evolutionary stages to within a few thousand years
\citep{2011ApJ...742..109E}. 

The millimetre wavelength maser transitions have simpler spectra than those at
lower frequency, suggesting that the high frequency lines come from more
compact gas clumps, however the parameter space probed by Band 2 is barely
tested by observations.  The milli-arcsec resolution available with VLBI will
constrain the physical conditions on AU size-scales at the typical distances
of massive young stellar objects (MYSOs) by revealing which lines co-propagate
as well as distinguishing between disc, outflow and infall regions
complementing VLBI observations of the 6.7\,GHz Class II methanol masers
\citep[e.g.][]{2009A&A...502..155B}.

Formaldehyde masers also arise from the vicinity of MYSOs but they are less
common and less well-understood than methanol masers but the cm-wave
transitions have been observed to arise from within a few arcseconds of
methanol masers. However, few existing observations have had sufficient
resolution to verify whether they co-propagate with the methanol masers, that
is whether they arise from exactly the same gas.  \citet{2010ApJ...717L.133A}
showed formaldehyde masers populating a gap in the distribution of methanol
masers, suggesting a likely separation of $\sim2000$\,AU.  Formaldehyde is
unusual in showing deep, anomalous absorption and there has been much debate
on whether this is directly associated with the maser mechanism
\citep[e.g.][]{2008ApJS..178..330A}.

Periodic or quasi-periodic flares with periods from a few weeks to many months
are a remarkable property of many methanol masers
\citep[e.g.][]{2012IAUS..287..108M} and at least one formaldehyde maser
\citep{2010ApJ...717L.133A}, but the mechanism(s) driving the variations are
the subject of much speculation.  Small time delays between different
components of the same source suggest that the cause is a disturbance
propagating at a speed intermediate between the speed of light and bulk gas
motion. This suggests that the variations are, for example, mediated by dust
heating. Monitoring multiple frequencies at high resolution will show what
changes in physical conditions are associated with flares.  Early results from
the Korean VLBI Network suggest that mm-wave methanol masers are also variable
but only very limited studies have been done to date.

The transitions of SiO in Band 2 (Table~\ref{tab:si}) can also be
masers. Almost all known examples of SiO masers arise in the envelopes of
evolved stars \citep{2012msa..book.....G} where VLBI observations of the
masers have been used to image and monitor the evolution of the stellar
outflows \citep[e.g.][]{2010MNRAS.406..395G} and in one case determine a high
precision parallax distance \citep{2014PASJ...66...38M}.  In addition, the
polarization of SiO maser spots has been used to trace the morphology of the
stellar magnetic field close to the stellar photosphere
\citep[e.g.][]{1997ApJ...481L.111K}. Predictions for the polarization of the
SiO transitions in the ALMA frequency range show that lines in Band 2 reaching
fractional linear polarizations of $\sim20$\%
\citep{2013A&A...551A..15P}. However these simulations do not include some
effects which could increase the degree of polarization, so, for example,
although the models suggest $\sim30$\% linear polarization in the SiO J=1--0
transition (which occurs in ALMA Band 1), observations have detected masers
spots in this transition with close to 100\% polarization
\citep[e.g.][]{2012A&A...538A.136A} while linear polarizations of up to 60\%
have been detected in the J=5-4 transitions
\citep{2011ApJ...728..149V}. Circularly polarized SiO maser emission has also
been detected, and although some maser spots can be up to $\sim40$\% polarized
towards the star TX Cam, the median value is $\sim3-5$\% (compared to a median
linear polarization of $\sim25$\%) \citep{1997ApJ...481L.111K}. For
comparison, current VLBI measurements have an estimated circular polarization
accuracy of $0.5-1$\% or better \citep{2011A&A...533A..26K}.

Formaldehyde is better known for the extragalactic masers detected at 4-5 GHz,
notably in Arp 220 \citep[e.g.][]{1993ApJ...415..140B}.  Absorption and
emission are seen and the maser pump is controversial
\citep[e.g.][]{2007IAUS..242..437B} since the mechanism suggested for
megamasers would not work on the scale of individual star-forming
regions. Emission from Band 2 transitions has been detected by single dishes
but VLBI is needed to investigate these maser components.

\subsubsection{Event Horizon}

\citet{2012Msngr.149...50F} summarise the meeting held to discuss VLBI with
ALMA and its potential to study the Event Horizon around a black hole.  Each
frequency penetrates a different depth into the black hole environment.  Full
frequency coverage is needed to distinguish between changes in spectral index
due to scattering or due to intrinsic changes in emitting material and
relativistic effects.  Although in Band 2 scattering will be a significant
issue for Sgr A*, a study of M87 using Band 2 would probe the material at
around $\sim12$ Schwarzchild radii where the spectral energy distribution
deviates from a power law and variability is increasingly rapidly
\citep[see][]{2012ApJ...760...52H}.

\subsubsection{Jet Launching and Shocks}

Each mm-VLBI frequency samples the jet-launching region around AGN to a unique
depth, probing the launching of the jet and the associated shocks.  These
phenomena can be resolved on milliarcsecond scales. Good frequency coverage,
including Band 2, is essential to trace the spectral curvature, identify the
jet flaring region and avoid ambiguities in Faraday rotation analysis due to
the polarization angle changing very rapidly with frequency
\citep{2011ApJ...737...42P}.  Resolved polarization images will tackle the
question of the role of magnetic fields in jet collimation, including Faraday
rotation mapping and controversial topics such as possible helicity.  The
polarization vector orientation constrains the shock model as well as the
evolution of the magnetic field after flares \citep{2013MNRAS.428.2418O} which
at millimetre wavelengths preceed gamma-ray flares suggesting that the
millimetre flares arise further from the nucleus and are possibly related to
superluminal ejections. Another controversy concerns whether the associated
shocks are due to ballistic ejecta or perturbations in a more continuous jet.

The millimetre-VLBI and gamma-ray behaviour of Mrk\,501 suggests, as-yet
unconfirmed, more distant jet limb-brightening/knot excitation. The spectrum
turns over at $\sim$15\,GHz, and the flux density halves between ALMA Bands 1
and 3 so Band 2 VLBI provides an ideal balance of resolution and flux density
\citep{2008A&A...488..905G}.  Examining the small-scale structure of compact
active galaxies also tackles questions such as the origins of FRI's and the
evolution of BL Lac objects (e.g. Liuzzo et al.\ in prep.) where the SEDs are
likely to show significant structure in the frequency range covered by Bands 2
and 3.

\subsubsection{High-z Absorption Kinematics, Chemistry and Fundamental
  Constants}

HI absorption has been mapped against the cores of a number of active
galaxies, providing important high resolution kinematic information
\citep[e.g.][]{1996IAUS..175...92C}. Similar CO and CII studies will allow
comparisons between the different components of the gas closest to the core.
Absorption in the low-J transitions of CO and other species can be imaged at
as good, or better, resolution as emission at the highest ALMA frequencies.

Previous millimetre absorption studies towards the lensed quasar PKS1830-211
have provided a spectacular molecular inventory of 42 species and 14 isotopomers
in a $z=0.89$ spiral Galaxy (ATCA survey, \citet{2011A&A...535A.103M}).  These
molecules can be used as probes of the gas physical conditions, in particular
to measure the temperature of the cosmic microwave background as a function of
redshift \citep{2013A&A...551A.109M}.  The presence of multiple transitions
from the same species, or from isotopomers, allows the investigation of
cosmological variations in the fundamental constants of nature, since the line
spacing is sensitive to variations in $\alpha$ (the fine structure constant) or
$\mu$ (the proton to electron mass ratio) \citep[e.g.][]{2013Sci...339...46B,
  2008Sci...320.1611M}.

The high angular resolution of VLBI is crucial in order to resolve molecular
absorption. Sub-parsec scale imaging is essential to ensure that the molecules
are cospatial \citep{2013ApJ...764..132S}, thus avoiding kinematic and
chemical biases in these analyses. Bands 2 and 3 are rich in low-energy
transitions of numerous species and are well-suited for absorption studies
(with bright continuum sources and high dynamic range achievable).  At such
small angular scales, the continuum emission is liable to be variable and the
availability of a wide band is advantageous to allow as many species as
possible to be observed within a short time interval.

\subsection{Cold Complex Chemistry}

Glycine (NH$_2$CH$_2$COOH) is the simplest of all amino acids. Its detection
in the interstellar medium is thus of key importance to understand the
formation mechanisms of pre-biotic molecules relevant to life and potentially
the origin of life on Earth. For over a decade there have been extensive
searches for glycine toward high-mass star forming regions such as the hot
molecular cores in SgrB2(N), Orion KL and W51 e1/e2
\citep[see][]{2003ApJ...593..848K}, although these studies have not yielded
any detection \citep{2005ApJ...619..914S, 2007MNRAS.376.1201C,
  2007MNRAS.374..579J}. Glycine and other amino acids have however been found
in meteorites (for example Wild 2; \citealt{2009M&PS...44.1323E}), consistent
with the idea that amino acids could have an interstellar origin
\citep{2000ARA&A..38..427E}.

Several challenges are faced in the search of glycine in high-mass star
forming regions. The spectral line densities measured toward hot molecular
cores are usually high, leading to a high level of line blending and thus, of
line confusion. This is due to the fact that hot cores show a very rich
chemistry in complex organic molecules driven by the thermal desorption of the
mantles of dust grains at the high gas/dust temperatures found in these
objects (100-200 K; \citealt[see e.g.][]{1999PASP..111.1049G}. The typical
linewidths of molecular rotational lines in hot cores are several km~s$^{-1}$,
which also prevents a clear identification of the glycine lines which are
expected to be weak. In addition, previous studies mainly targeted glycine
lines at frequencies >100 GHz because the energies of their upper levels
($E_u$) were $>70$~K, i.e. similar to the temperatures found in hot cores
\citep[see e.g.][]{2005ApJ...619..914S}. And for the transitions at lower
frequencies (such as those in the 3mm band, between 85 and 105\,GHz), previous
interferometric observations were not sensitive enough and only provided upper
limits to the total column density of glycine toward the SgrB2 hot molecular
core (of $10^{14} -10^{15}$ cm$^{-2}$; \citealt{2007MNRAS.374..579J}).

\subsubsection{Glycine in Low-mass Star Forming Regions}

In contrast with their high-mass counterparts, low-mass star forming regions
may be better suited for the detection of glycine in the interstellar medium
for several reasons. The level of line confusion is expected to be low in
low-mass star forming regions (especially at their early stages represented by
pre-stellar cores) because the measured gas temperatures are $<10$~K and thus
the number of molecular rotational lines that can be excited at these
temperatures is significantly smaller. In addition, the typical linewidths in
low-mass cores are $<0.5$ km~s$^{-1}$, which allows a more accurate
identification of the observed transitions since the lines likely suffer less
from line blending. Complex organic molecules such as propylene
(CH$_2$CHCH$_3$) have indeed been detected toward molecular dark cores such as
TMC-1 \citep{2007ApJ...665L.127M}, suggesting that non-canonical chemical
mechanisms could be playing a key role in the formation of these large
molecules in the interstellar medium \citep{2013MNRAS.430..264R}.


\begin{figure}[ht!]
\includegraphics[width=8cm,origin=c,angle=-90]{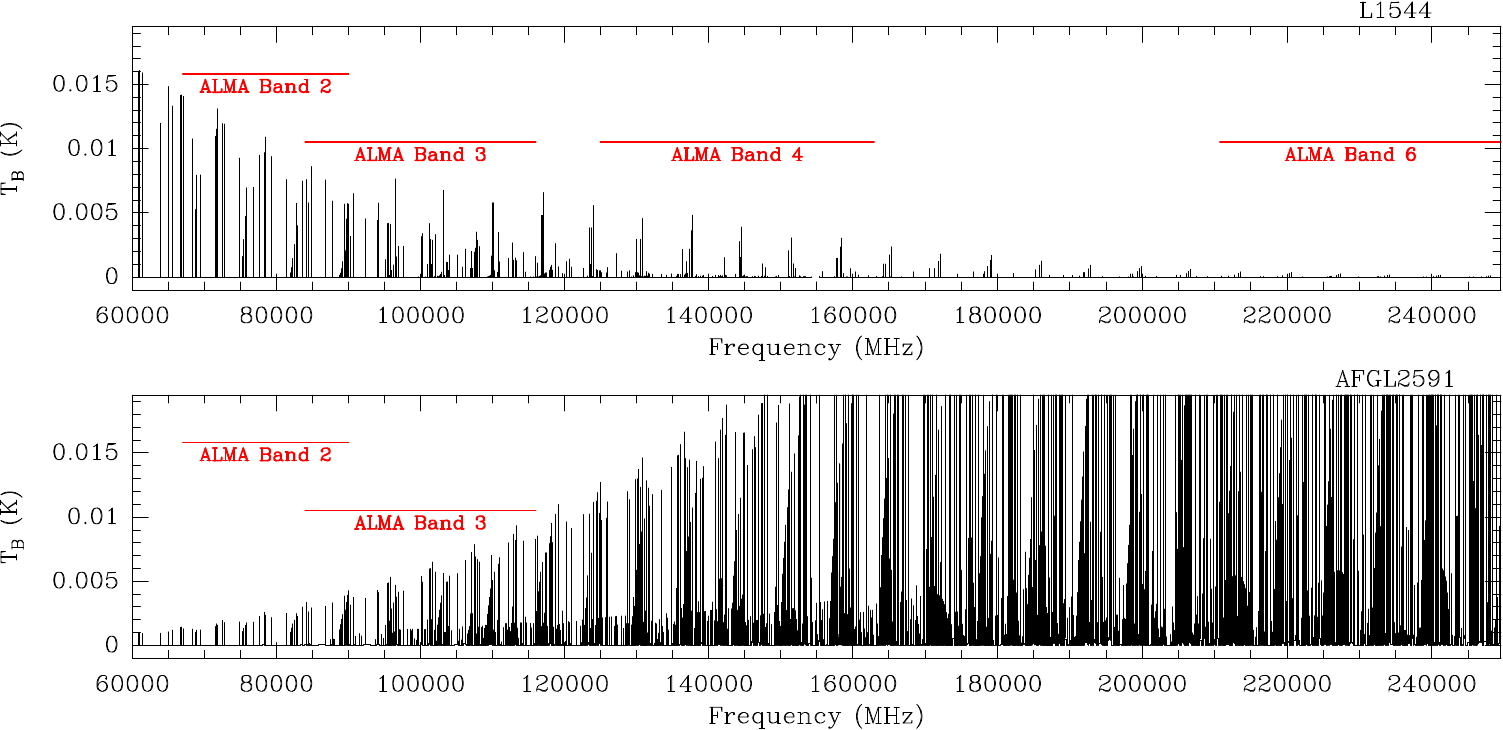}
\vspace{-4cm}
  \caption{\footnotesize Upper panel: LTE predictions of the line intensity
    for the glycine transitions between 60\,GHz and 250\,GHz towards the
    pre-stellar core L1544. Lower panel: LTE predictions for the same lines
    towards the massive hot core in AFGL2591.}
\label{fig:glycine}

\end{figure}


The high-sensitivity of ALMA opens up the possibility to detect a large number
of complex organic molecules of biochemical interest such as glycine in
low-mass star forming regions. ALMA has therefore the potential to provide new
insight into the formation processes of pre-biotic molecules in Space and into
their subsequent delivery onto planetary systems. Low-mass star forming
objects at different stages in their evolution (such as the pre-stellar core
L1554, the low-mass warm core (or hot corino) IRAS16293, and the
protoplanetary disk TW Hya) should be targeted so that the different formation
routes of glycine can be analysed in detail along the formation process of
low-mass stars. As shown below, ALMA Band 2 is very well suited for the
discovery of glycine in young sun-like systems, because it covers the lowest
excitation transitions of this molecule with the highest possible Einstein A
coefficients.

Most of the glycine transitions observed in previous studies toward hot
molecular cores are found in the 1.3mm wavelength range
\citep{2005ApJ...619..914S}. This is due to the fact that their Einstein A
coefficients are $\sim10^{-5}$ s$^{-1}$, one order of magnitude higher than
those of the 3mm lines. However, the values of the upper state energys, $E_u$,
for these transitions are $>200$~K. In the frequency range between 65 and 116
GHz (i.e. Band 2+3), glycine (conformer I) has several rotational transitions
whose energy levels lie below 25~K and whose Einstein A coefficients are
$\sim10^{-6}$ s$^{-1}$.  The lowest frequency transition of glycine with an
Einstein A coefficient $>10^{-6}$ s$^{-1}$ is indeed located at 60873.034 MHz.
For frequencies lower than 72~GHz, there are 10 glycine lines with $E_u<25$~K
and an Einstein A coefficient $>10^{-6}$ s$^{-1}$,
with a further 48 lines below 90~GHz with Einstein A coefficient $>10^{-6}$
s$^{-1}$ and $E_u< 63$~K of which 37 have $E_u<40$~K.  Note that we only refer
to glycine conformer I because the ground vibrational level of glycine
conformer II lies 700 cm$^{-1}$ (i.e. $\sim1000$~K) above that of the
conformer I.



\begin{wraptable}{r}{11cm}
\caption{\small Sample of the brightest predicted glycine lines from L1544}
\label{tab:glycine}
\centering
\begin{tabular}{lcccc}
\toprule
Line & Transition & Frequency$^a$ & Einstein A & Energy \\
&&  (MHz)   & ($10^{-6}$ s$^{-1}$)   & (K)   \\
\midrule
1 &$10_{1,9}-9_{1,8}$ &67189.12 & 1.3  & 18.8 \\
2 &$10_{3,8}-9_{3,7}$ &68323.70 &  1.3 &  21.1\\
3 &$10_{3,7}-9_{3,6}$ &71611.56 &  1.5 &  21.5\\
4 &$11_{2,10}-10_{2,9}$ &71646.39 &  1.6 &  22.4\\
5 &$10_{2,8}-9_{2,7}$ &71910.30 & 1.6 &  20.2\\
6 &$12_{1,12}-11_{1,11}$ &72559.35 & 1.7 & 23.3 \\
7 &$12_{0,12}-11_{0,11}$ &72061.11 & 1.7 & 23.3 \\
8 &$11_{1,10}-10_{1,9}$ &72841.25 & 1.7 & 22.3 \\
9 &$11_{2,9}-10_{2,8}$ &78524.63 & 2.1 & 24.0 \\
 \bottomrule
\end{tabular}

\noindent $^a$ \citet{2005JMoSt.742..215M}
\end{wraptable}


Figure~\ref{fig:glycine} (upper panel) shows the LTE spectrum of 
glycine between 60 and 250\,GHz towards the L1544 pre-stellar core. 
The spherically symmetric radiative transfer calculations used the
physical structure of the L1544 core derived by \citet{2010MNRAS.402.1625K}
and the gas-phase water abundance profile inferred by
\citet{2012ApJ...759L..37C} toward this core. Glycine is assumed to desorb
together with water, following its gas-phase abundance profile but scaled down
by the fraction of glycine present in the ices (solid abundance of glycine
$\sim10^{-4}$ with respect to water;
\citealt{2002Natur.416..401B,2002Natur.416..403M}). This translates into
gas-phase abundances of $\sim3\times10^{-11}$ for glycine in L1544
\citep{2014ApJ...787L..33J}. The glycine emission is expected to arise from
the innermost few thousand AU of L1544 corresponding to angular scales of
$\sim10$''. These scales can be easily imaged by ALMA in Band 2 in its compact
configuration.

Figure~\ref{fig:glycine} (upper panel) shows that several glycine lines in
Band 2+3 have peak intensities $>5$\,mK. In particular, the brightest lines
with intensities $>8$\,mK are found below 80 \,GHz (Table~\ref{tab:glycine}),
making these transitions prime targets for ALMA Band 2. Since the molecular
gas in L1544 is very cold, the higher energy levels of glycine are not
populated efficiently, making the glycine lines in Band 4 and Band 6 very
weak. The transitions below 80\,GHz are therefore key to probing cold glycine
at the early stages in the formation of low-mass stars.

For comparison, Figure~\ref{fig:glycine} (lower panel) shows the LTE
predicted spectrum of glycine from a massive, hot molecular core. For
these calculations the physical structure of the AFGL2591 hot core derived by
\citet{2012ApJ...753...34J} was adopted. It was assumed that glycine is
evaporated fully from ices and therefore its abundance in the gas phase is
$\sim10^{-8}$ \citep{2014ApJ...787L..33J}.

Despite the larger abundance of glycine in hot sources, its transitions in
Band 2+3 are factors of $>3-15$ weaker in hot sources than in cold
objects. This is mainly due to the lower efficiency when populating the low
energy levels of glycine at temperatures $>100$\,K. Although the glycine lines
in Band 4 and Band 6 are bright, line blending and line confusion present
problems for the detection and identification of low-abundant species such as
amino acids. These models show that with ALMA Band 2+3 receivers the discovery
of glycine would be possible toward the coldest phases of the formation of
Solar-type systems.

\subsection{Solar System Science: Time Variability}

\begin{figure}
\centering
\includegraphics[width=.5\columnwidth]{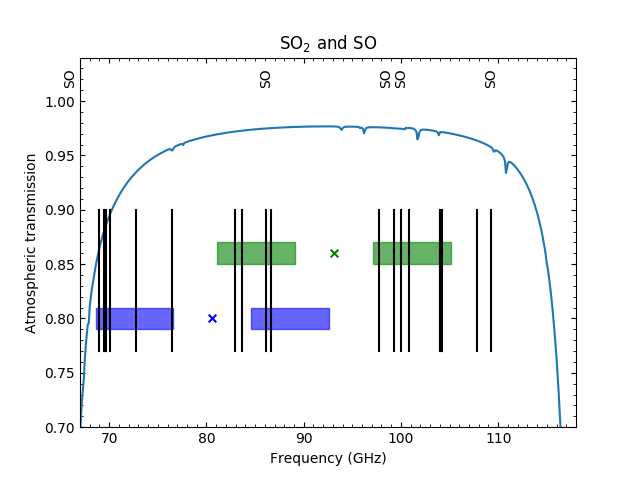}\includegraphics[width=.5\columnwidth]{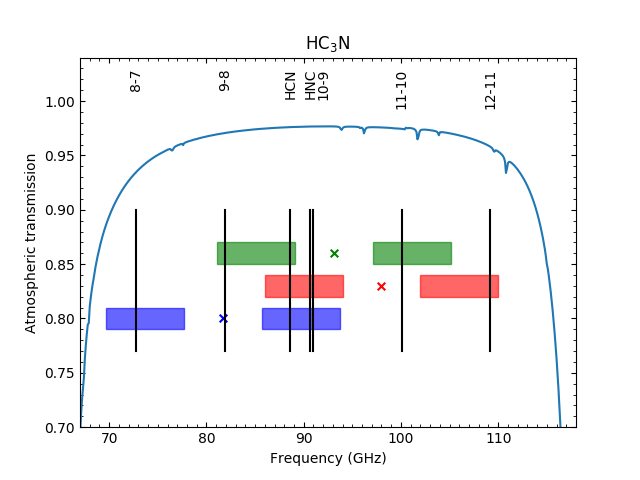}
\caption{The pairs of coloured horizontal bars indicate the frequency coverage of the upper and lower sidebands of Band 2+3 with 8\,GHz bandwidth per sideband with the LO frequency marked by the coloured cross midway between the bars. The vertical lines mark the frequencies of molecular transitions and the blue curve, the zenith atmospheric transmission at the ALMA site for 2mm pwv.  {\it Left:} The black vertical lines show a range of SO and SO$_2$ transitions. Those from SO are indicated  and the remainder of the lines are from SO$_2$. Two LO settings are shown which simultaneously cover a range of transitions of both species. {\it Right:} The black vertical lines show the J=8-7, 9-8, 10-9, 11-10 and 12-11 transitions of HC$_3$N  and the J=1-0 transitions of HCN and HNC. Three different possible LO settings are shown. Each setting covers two transitions of HC$_3$N. Two of these settings also over both HCN and HNC while the other covers only HCN. Each of these settings will also cover a range of HC$_3$N transitions in  vibrationally excited levels. }
\label{fig:so-hc3n}
\end{figure}

An advantage of a broad instantaneous bandwidth is the ability to simultaneously observe transitions which vary in time such as from comets and other solar system bodies. For example, studying the SO and SO$_2$ on Io where outgassing from volcanoes and the sublimation of SO$_{\rm 2}$ frost drive the composition of the atmosphere \citep{2014DPS....4641101M}. Or probing the structure of the hydrocarbon rich atmosphere of Titan where HNC and HC$_3$N trace different ranges of heights in the atmosphere \citep{2014ApJ...795L..30C,2018ApJ...859L..15C}. As Figure~\ref{fig:so-hc3n} shows, Band 2+3 with 8\,GHz per sideband enables the simultaneous observation of a range of SO and SO$_2$ transitions as well as allowing the simultaneous observation of HNC J=1-0 and two different J transitions of HC$_3$N. Simultaneous observations of multiple transitions of a species provide much better constraints on the varying excitation and abundance of a molecule than observations of single transitions.  The wide range of species detected in comets probe the composition and history of some of the more primitive material in our solar system and recently, other, solar systems \citep{2018ASPC..517...73C,2020NatAs.tmp...84C}. Their highly variable emission makes simultaneous observations of related species in comets essential for reliable relative abundance determinations to, for example, measure the deuteration fraction with its implication for the transport of volatile species in the early solar system. The value of Band 2+3 for deuterium measurements is discussed above in Section~\ref{sec:deuterium}.

\bibliography{band2}

\newpage
\theendnotes

\end{document}